\documentclass[twocolumn]{aastex63}

\usepackage{epstopdf}
\usepackage{mathtools}
\usepackage{mathrsfs}
\usepackage{calrsfs}
\usepackage{multirow}
\usepackage{xspace}
\usepackage{xcolor}
\usepackage[hang,flushmargin]{footmisc}
\usepackage[autostyle, english = american]{csquotes}
\MakeOuterQuote{"}
\usepackage{eucal}
\usepackage{amsmath}

\usepackage{bm}

\newcommand{\V}{{\bm V}}
\newcommand{\D}{{\bm D}}
\newcommand{\G}{{\bm G}}
\newcommand{\I}{{\bm I}}
\newcommand{\R}{{\bm R}}
\newcommand{\Par}{{\bm P}}
\newcommand{\J}{{\bm J}}

\newcommand{\pone}{Paper I}

\received{--}
\revised{--}
\accepted{--}

\submitjournal{ApJ}

\shortauthors{Park et al.}

\graphicspath{{./}{}}

\begin{document}

\title{Calibrating VLBI Polarization Data Using GPCAL. II. Time-Dependent Calibration}

\correspondingauthor{Jongho Park}
\email{jpark@kasi.re.kr}

\author[0000-0001-6558-9053]{Jongho Park}
\affiliation{Department of Astronomy and Space Science, Kyung Hee University, 1732, Deogyeong-daero, Giheung-gu, Yongin-si, Gyeonggi-do 17104, Republic of Korea}

\affiliation{Korea Astronomy and Space Science Institute, Daedeok-daero 776, Yuseong-gu, Daejeon 34055, Republic of Korea}
\affiliation{Institute of Astronomy and Astrophysics, Academia Sinica, P.O. Box 23-141, Taipei 10617, Taiwan}

\author[0000-0001-6988-8763]{Keiichi Asada}
\affiliation{Institute of Astronomy and Astrophysics, Academia Sinica, P.O. Box 23-141, Taipei 10617, Taiwan}

\author[0000-0003-1157-4109]{Do-Young Byun}
\affiliation{Korea Astronomy and Space Science Institute, Daedeok-daero 776, Yuseong-gu, Daejeon 34055, Republic of Korea}
\affiliation{University of Science and Technology, Gajeong-ro 217, Yuseong-gu, Daejeon 34113, Republic of Korea}

\begin{abstract}

We present a new method of time-dependent instrumental polarization calibration for Very Long Baseline Interferometry (VLBI). This method has been implemented in the recently developed polarization calibration pipeline GPCAL. Instrumental polarization, also known as polarimetric leakage, is a direction-dependent effect, and it is not constant across the beam of a telescope. Antenna pointing model accuracy is usually dependent on time, resulting in off-axis polarimetric leakages that can vary with time. The method is designed to correct for the off-axis leakages with large amplitudes that can severely degrade linear polarization images. Using synthetic data generated based on real Very Long Baseline Array (VLBA) data observed at 43 GHz, we evaluate the performance of the method. The method was able to reproduce the off-axis leakages assumed in the synthetic data, particularly those with large amplitudes. The method has been applied to two sets of real VLBA data and the derived off-axis leakages show very similar trends over time for pairs of nearby sources. Furthermore, the amplitudes of the off-axis leakages are strongly correlated with the antenna gain correction factors. The results demonstrate that the method is capable of correcting for the off-axis leakages present in VLBI data. By calibrating time-dependent instrumental polarization, the rms-noise levels of the updated linear polarization images have been significantly reduced. The method is expected to substantially enhance the quality of linear polarization images obtained from existing and future VLBI observations.

\end{abstract}

\keywords{high angular resolution --- techniques: interferometric --- techniques: polarimetric --- methods: data analysis}

\section{Introduction} \label{sec:intro}

Polarization observations using Very Long Baseline Interferometry (VLBI) are ideally suited for studying the processes of mass accretion and jet formation (see, e.g., \citealt{PA2022} and references therein). These processes occur at small physical scales and can only be observed with instruments with high angular resolution. Polarized emission from plasma in mass accretion flows and jets is directly related to their magnetic fields. Recent observations of the nearby elliptical galaxy M87 with the Event Horizon Telescope (EHT) demonstrate the power of VLBI polarimetry \citep{EHT2019a, EHT2019b, EHT2019c, EHT2019d, EHT2019e, EHT2019f}. The EHT total intensity image of M87 reveals the presence of a prominent ring-like structure. This structure is interpreted as the result of gravitational light bending of synchrotron emission from a hot plasma surrounding the black hole, along with photon capture occurring at the event horizon. In the corresponding linear polarization images of the ring, the polarization position angles are arranged in a nearly azimuthal pattern \citep{EHT2021a}. The comparison between these images and the General Relativistic Magnetohydrodynamic (GRMHD) simulation results indicates that a strong ordered, poloidal-dominated magnetic field may exist around the black hole, which is capable of generating the powerful jets seen in this source. \citep{EHT2021b}. In addition, polarization observations of AGN jets using VLBI have provided important information about the collimation, acceleration, and particle acceleration processes in AGN jets \citep[e.g.,][]{Asada2002, Jorstad2017, Gabuzda2018, Lister2018, Park2019a, Park2021c, Lisakov2021}.

The linear polarization of AGN jets typically ranges from a few to a few tens of percent \citep[e.g.,][]{Jorstad2017, Lister2018}. In the past, the quality of VLBI linear polarization images of AGN jets was determined primarily by their sensitivity. However, recent VLBI arrays have increased their sensitivity considerably by enlarging the recording bandwidth and including very high sensitivity stations such as the phased Atacama Large Millimeter/submillimeter Array (ALMA; \citealt{Matthews2018}) and the phased Very Large Array (VLA). Consequently, systematic errors are becoming a more critical factor in determining the quality of VLBI linear polarization images. The most dominant systematic error affecting VLBI polarization data is "instrumental polarization". The source of this polarization signal is due to the antenna polarization not being exactly circular or linear, causing an unpolarized source to appear polarized. This polarization is commonly referred to as "polarization leakage" or "D-terms" and it must be corrected in the visibility data before producing an image.

LPCAL is a task incorporated into Astronomical Image Processing System \citep[AIPS,][]{Greisen2003} and based on the linearized leakage model \citep{Leppanen1995}. It has been a standard program for calibrating instrumental polarization in VLBI data for a long time. In spite of its success for many studies utilizing various VLBI arrays \citep[e.g.,][]{Casadio2017, Jorstad2017, Lister2018, Park2018, Park2019a, Gomez2022, Zhao2022}, there are some limitations which prevent accurate calibration. An important limitation is that the "similarity approximation"\footnote{As far as we are aware, the term "similarity approximation" was initially introduced in \cite{Leppanen1995}. The concept was originally proposed by \cite{Cotton1993}, who also noted that the assumption of a direct proportionality between total intensity and linear polarization emission may not be valid.}, which assumes similar total intensity and linear polarization structures for the calibrators \citep{Cotton1993, Leppanen1995}, is often violated for VLBI. The problem becomes more severe at high frequencies, where nearly all calibrators are resolved. Recently, a number of new calibration and imaging pipelines have been developed to improve calibration accuracy and linear polarization images, including the Generalized Polarization Calibration pipeline (GPCAL; \citealt{Park2021a}), \texttt{polsolve} \citep{MartiVidal2021}, the \texttt{eht-imaging} software library \citep{Chael2016, Chael2018}, D-term Modeling Code (DMC, \citealt{Pesce2021}), THEMIS \citep{Broderick2020}, and Comrade \citep{Tiede2022}. Some of the pipelines were applied to the first linear polarization imaging of the M87 black hole \citep{EHT2021a}.

Nevertheless, most of the existing pipelines rely on the fundamental assumptions that instrumental polarization remains constant over a wide range of frequencies of a receiver and over a period of time during observation. VLBI arrays of recent generations can violate these assumptions, which results in systematic errors in the visibility data. This series of papers presents new methods for modeling the frequency- and time-dependent instrumental polarization in VLBI data, which have been implemented in GPCAL. The method for correcting frequency-dependent instrumental polarization is presented in a companion paper (Park et al. 2023; henceforth \pone{}). The purpose of this paper is to introduce the method for correcting for time-dependent instrumental polarization. 

Instrumental polarization itself is believed to not change significantly during the observation. However, it is a direction-dependent effect \citep[e.g.,][]{Smirnov2011}, so the instrumental polarization varies depending on the direction of the antenna beam (due to the cross polarized sidelobes; see, e.g., \citealt{Napier1989, Thum2008}). More specifically, instrumental polarization of an antenna consists of two main components (see Section 4.3 in \citealt{Napier1989}). There is one at the center of the antenna beam that can be considered constant across the beam ($D_{\rm on-axis}$). The other component varies across the beam ($D_{\rm off-axis}$). The accuracy of antenna pointing during observation can be subject to variations over time, arising from stochastic winds and the deformation of antennas caused by sunlight, among other factors. Thus, effective D-terms can vary over time during an observation due to the direction-dependent effect. Only if the antenna pointing is sufficiently accurate throughout the observing period does instrumental polarization remain constant. Time-dependent leakage amplitudes are expected to be directly related to antenna pointing accuracy in this case. In general, high-frequency observations and antennas with large diameters usually have less accurate antenna pointing due to small beam sizes and stronger dish deformation at low elevations. Moreover, the instrumental polarization of phased arrays (such as the phased ALMA and VLA) is variable \citep[e.g.,][]{EHT2019c}, as is the phasing efficiency. As demonstrated in this paper, even the D-terms for the Very Long Baseline Array (VLBA) may vary greatly due to inaccurate antenna pointing and changing weather conditions. GPCAL can increase the dynamic range of the VLBA linear polarization images by a factor of several by correcting the time-dependent instrumental polarization.

The paper is organized as follows. In Section~\ref{sec:model}, we describe the radio interferometer measurement equation, which forms the basis of the GPCAL instrumental polarization model. We discuss in Section~\ref{sec:pipeline} the strategy for calibrating time-dependent polarization leakages. We validate the performance of the method with a synthetic data set in Section~\ref{sec:synthetic}. In Section~\ref{sec:application}, we apply the method to real VLBA data sets and verify that time-dependent leakage correction using GPCAL can significantly enhance the quality of VLBI linear polarization images. We summarize and conclude in Section~\ref{sec:conclusion}.

\section{The Radio Interferometer Measurement Equation} 
\label{sec:model}

We use the radio interferometer measurement equation (RIME; \citealt{Hamaker1996, Sault1996, HB1996, Hamaker2000, Smirnov2011}), as described in Section 2 in \pone{}. In this paper, we provide a brief explanation of the equation for the convenience of readers.

The observed complex visibilities between two VLBI antennas, $m$ and $n$, are expressed in a visibility matrix, $\V_{mn}$, of the form
\begin{equation}
\V_{mn} = 
\begin{pmatrix}
r^{RR}_{mn} & r^{RL}_{mn}\\
r^{LR}_{mn} & r^{LL}_{mn}
\end{pmatrix},
\end{equation}
where $R$ and $L$ refer to the right- and left-handed circular polarizations (RCP and LCP), respectively. The observed $V_{mn}$ is corrupted by antenna gain and polarization leakage. It is convenient to arrange all the corruptions into a single Jones matrix \citep{Jones1941}:
\begin{eqnarray}
    \J_{m} &=& \G_m \D_m \Par_m \nonumber\\
    &=& 
    \begin{pmatrix}
    G^R_m & 0 \\
    0 & G^L_m
    \end{pmatrix}
    \begin{pmatrix}
    1 & D^R_m \\
    D^L_m & 1
    \end{pmatrix}
    \begin{pmatrix}
    e^{j\phi_m} & 0 \\
    0 & e^{-j\phi_m}
    \end{pmatrix},
\end{eqnarray}
where $G$ is the complex antenna gain, $D$ is the leakage factor (D-term), and $\phi$ is the antenna field rotation angle. Subscripts  and superscripts denote antenna numbers and polarization, respectively. The field rotation angle is a combination of the source's elevation ($\theta_{\rm el}$), parallactic angle ($\psi_{\rm par}$), and a constant offset for the rotation of antenna feed with respect to the azimuth axis ($\phi_{\rm off}$) via:
\begin{equation}
    \phi = f_{\rm el}\theta_{\rm el} + f_{\rm par}\psi_{\rm par} + \phi_{\rm off},
\end{equation}
Cassegrain mounts have $f_{\rm par} = 1$ and $f_{\rm el} = 0$. Nasmyth-Right type mounts have $f_{\rm par} = 1$ and $f_{\rm el} = +1$ and Nasmyth-Left type mounts have $f_{\rm par} = 1$ and $f_{\rm el} = -1$.

The observed $\V_{mn}$ are modifications of the true $\bar\V_{mn}$ through:
\begin{equation}
\label{eq:rime}
    \V_{mn} = \J_m \bar\V_{mn} \J^H_n,
\end{equation}
where H is the Hermitian operator.

For circular feeds, the $\bar\V$ are related to the Fourier transforms of the Stokes parameters ($\tilde{I}$, $\tilde{Q}$, $\tilde{U}$, and $\tilde{V}$) via
\begin{equation}
\label{eq:stokes}
\bar\V_{mn} \equiv
\begin{pmatrix}
\mathscr{RR} & \mathscr{RL} \\
\mathscr{LR} & \mathscr{LL}
\end{pmatrix}
=
\begin{pmatrix}
\tilde{I}_{mn} + \tilde{V}_{mn} & \tilde{Q}_{mn} + j\tilde{U}_{mn} \\
\tilde{Q}_{mn} - j\tilde{U}_{mn} & \tilde{I}_{mn} - \tilde{V}_{mn}
\end{pmatrix}.
\end{equation}

As explained in Paper I, GPCAL assumes that the antenna field-rotation angles are already corrected at an upstream calibration stage (before performing global fringe fitting). Thus, Equation~\ref{eq:rime} becomes:
\begin{equation}
\label{eq:matmodel}
    \V_{mn} = \Par^{-1}_m \G_m \D_m \Par_m \bar\V_{mn} \Par^H_n \D^H_n \G^H_n (\Par^H_n)^{-1},
\end{equation}

\section{Calibration Procedure}
\label{sec:pipeline}

\subsection{Model Equation}

GPCAL assumes that antenna gains are already corrected during the upstream calibration and imaging/self-calibration procedures. The original GPCAL pipeline \citep{Park2021a} fits Equation~\ref{eq:matmodel} with the assumption of $\G = \I$ to the observed cross-hand visibilities averaged over the frequency bandwidth. The pipeline assumes that polarimetric leakages are constant during the observation. However, if this assumption is violated, there are residual leakages in the data. Following the Appendix in \cite{Pesce2021}, we can write the true leakage matrix $\D$ as:
\begin{equation}
    \D =  
    \begin{pmatrix}
    1 & D^R \\
    D^L & 1
    \end{pmatrix},
\end{equation}
and the estimated leakage matrix $\hat\D$ with the assumption of constant leakage terms during the observation, which can be different from $\D$, as:
\begin{equation}
    \hat{\D} =  
    \begin{pmatrix}
    1 & D^R + \Delta^R \\
    D^L + \Delta^L & 1
    \end{pmatrix}.
\end{equation}

The visibility matrix after running the GPCAL pipeline would become:
\begin{eqnarray}
\label{eq:resmodel}
    \V_{mn} &=& [\Par^{-1}_m \hat\D_m \Par_m]^{-1}[\Par^{-1}_m \D_m \Par_m] \bar\V_{mn} \nonumber\\ 
    &&[\Par^H_n \D^H_n (\Par^H_n)^{-1}][\Par^H_n \hat\D^H_n (\Par^H_n)^{-1}]^{-1} \nonumber\\
    &=&\Par^{-1}_m \R_m \Par_m \bar\V_{mn} \Par^H_n \R^H_n (\Par^H_n)^{-1},
\end{eqnarray}

where $\R$ is a residual leakage matrix:
\begin{eqnarray}
    \R &\equiv& \hat\D^{-1} \D \nonumber\\ 
    &\approx&
    \begin{pmatrix}
    1 & -\Delta^R \\
    - \Delta^L & 1
    \end{pmatrix}.
\end{eqnarray}

The approximation holds when dropping out second-order terms. With this approximation, the off-diagonal terms in $\R$ are the "residual" leakage terms, i.e., $[\R_{i,j} = \D_{i,j} - \hat\D_{i,j}]_{\in i\ne j}$.

First, we use the GPCAL pipeline to remove polarimetric leakages that are assumed to remain constant over time ($\hat\D$). We then fit Equation~\ref{eq:resmodel} to the corrected data in order to derive the "residual" time-dependent leakages. According to \cite{Pesce2021}, it may be more appropriate to redo calibration entirely with raw data (after data pre-processing) rather than incrementally calibrating a partially calibrated data set. In spite of this, we use this two-step procedure because, as we will demonstrate below, the signal-to-noise ratio of data plays a crucial role in the accurate estimation of time-dependent leakages. Therefore, before performing time-dependent instrumental polarization calibration, it is preferable to average the data over frequency. Nevertheless, as demonstrated in \pone{}, modern VLBI arrays provide a large recording bandwidth, which can lead to significant variations in the D-terms over frequency. The frequency-dependent D-terms may introduce non-negligible non-closing errors in the data if they are not removed before averaging the data over frequency. It is possible to accurately correct frequency-dependent D-terms using our method presented in \pone{} or even the GPCAL pipeline, which uses averaged data within each IF, which can remove the gross variations of D-terms across the entire frequency band. Thus, we use the data after correcting for $\hat\D$, where the gross variations of D-terms over frequency have already been removed, and correct for the residual time-dependent leakages.

The cross-hand visibilities in Equation~\ref{eq:resmodel} can be re-written as:
\begin{eqnarray}
\label{eq:fitmodel}
r^{RL}_{mn} &\approx& (\tilde{Q}_{mn} + j\tilde{U}_{mn}) + \Delta^R_m (t) e^{2j\phi_m}r^{LL}_{mn, {\rm cal}} \nonumber \\ &+& \Delta^{L*}_n(t)e^{2j\phi_n}r^{RR}_{mn, {\rm cal}} \nonumber \\ &+& \Delta^R_{m}(t)\Delta^{L*}_{n}(t)e^{2j(\phi_m+\phi_n)}(\tilde{Q}_{mn} - j\tilde{U}_{mn}) \nonumber \\
r^{LR}_{mn} &\approx& (\tilde{Q}_{mn} - j\tilde{U}_{mn}) + \Delta^L_m (t) e^{-2j\phi_m}r^{RR}_{mn, {\rm cal}} \nonumber\\ &+& \Delta^{R*}_n (t) e^{-2j\phi_n}r^{LL}_{mn, {\rm cal}} \nonumber\\ &+& \Delta^L_{m} (t) \Delta^{R*}_{n} (t) e^{-2j(\phi_m+\phi_n)}(\tilde{Q}_{mn} + j\tilde{U}_{mn}), \nonumber\\
\end{eqnarray}
where we replaced $\mathscr{RL}$ and $\mathscr{LR}$ using Equation~\ref{eq:stokes} and $\mathscr{RR}$ and $\mathscr{LL}$ by the final calibrated parallel-hand visibilities $r^{RR}_{mn, {\rm cal}}$ and $r^{LL}_{mn, {\rm cal}}$, respectively. We assume that $\tilde{Q}$ and $\tilde{U}$ are constant over time during the observation and depend only on the baseline coordinate $(u,v)$. We fit Equation~\ref{eq:fitmodel} to the data for each scan to derive $\Delta^R(t)$ and $\Delta^L(t)$ using the Scipy curve\_fit package.

\subsection{Calibration Strategy}

We fit Equation~\ref{eq:fitmodel} to the data for each scan. There is a limited number of data points used for fitting in this case. In a scan with $N$ antennas, there are $4N$ free parameters (the real and imaginary parts of the D-terms for RCP and LCP for each antenna). If one tries to fit a model with $4N$ degrees of freedom to a limited number of data points within a scan as compared to the entire dataset, there is likely to be a significant correlation among the parameters. We will demonstrate below that the best-fit D-terms in this case have large amplitudes because of the high correlation between parameters, while the assumed D-terms have small amplitudes in the synthetic data (Section~\ref{sec:synthetic}). This is similar to performing an amplitude self-calibration at a very short solution interval on the total intensity data of weak sources, which results in antenna gain solutions with very high amplitudes.

Due to the limited signal-to-noise ratio of the data (primarily as a result of the limited number of total data points) compared to the number of free parameters, it is challenging to accurately constrain the polarimetric leakages of all stations for each scan. Additionally, antenna pointing should be reasonably accurate for most stations and most scans. Nevertheless, some stations may have more inaccurate antenna pointing than others due to poor weather conditions or large diameters. For stations with inaccurate pointing during particular scans, the linear polarization models produced after correcting for on-axis instrumental polarization may differ substantially from the cross-hand visibilities of all baselines associated with those stations. If this is the case, one can assume that there are significant residual leakages ($\Delta$ in Equation~\ref{eq:fitmodel}) only for those stations for the scans, and fix the residual D-terms for the other stations to be zero for fitting. By doing so, we will be able to avoid the strong correlation between fitting parameters and improve the accuracy of the fitting\footnote{For fitting, the method uses the visibility weights that are stored in the UVFITS files provided by the users. However, the users have the option of scaling down the visibility weights of particular antennas by a constant factor, as was done in the original GPCAL pipeline \citep{Park2021a}. This feature proves advantageous for arrays featuring a large variance in sensitivity among the antennas, as it serves to prevent the fitting process from being dominated by the most sensitive stations.}. 

The first polarization imaging of M87 from the 2017 EHT observations \citep{EHT2021a} was also carried out using a similar approach. In some calibration pipelines, short baselines are used to calibrate the D-terms of the stations comprising the short baselines (ALMA and the Atacama Pathfinder Experiment (APEX) telescope in Chile and the James Clerk Maxwell Telescope (JCMT) and Submillimeter Array (SMA) on Maunakea in Hawaii). Those pipelines then assume that the D-terms of those stations have already been corrected and perform fitting for only the D-terms of the other stations based on the long baseline data. Through this two-step calibration strategy, which takes advantage of the fact that short baselines are less sensitive to complex source structures and have a high signal-to-noise ratio, it was possible to achieve good calibration accuracy and avoid strong correlations between the D-terms of many stations.

Accordingly, we adopt the following calibration strategy for time-dependent calibration of instrumental polarization.
\begin{enumerate}
    \item By using CLEAN with Difmap on the data corrected for the on-axis D-terms, Stokes $Q$ and $U$ images are produced. 
    \item For each scan and for each station $m$, we compute a norm
    \begin{equation}
        \chi^2_m = \sum_{n,k} w_{mn, k} \left(|r^{RL}_{mn, k} - \bar{r}^{RL}_{mn, k}|^2 + |r^{LR}_{mn, k} - \bar{r}^{LR}_{mn, k}|^2 \right),
    \end{equation}
    where $w_{mn, k}$ is the weight of the $k$th visibility matrix $V^k$ in the scan for the baseline between stations $m$ and $n$, $\bar{r}^{RL}_{mn, k} \equiv \tilde{Q}_{mn, k} + j\tilde{U}_{mn, k}$ and $\bar{r}^{LR}_{mn, k} \equiv \tilde{Q}_{mn, k} - j\tilde{U}_{mn, k}$ are the model cross-hand visibilities corresponding to the $k$th visibility for the baseline derived from the Fourier Transforms of the Stokes $Q$ and $U$ CLEAN images.
    \item Identify $l$ number of stations having the largest norm values (i.e., the stations giving the largest and the second largest $\chi_m^2$ for $l=2$). $l$ is a free parameter controlled by the variable \texttt{timecal\_freepar} in GPCAL, which is provided by users. The determination of the optimal value for \texttt{timecal\_freepar} may vary depending on the data and the SNR of the source, as indicated in Table 1 and Section 4. In the case of VLBA data, it is recommended to adopt a conservative approach and set a small value, such as \texttt{timecal\_freepar} = 1--2.
    \item We fit Equation~\ref{eq:fitmodel} to the cross-hand visibilities in the scan. Let the residual D-terms ($\Delta^R$ and $\Delta^L$) for the stations identified in Step 3 be free parameters and those for the other stations be fixed to be zero during the fitting.
    \item Repeat steps 1--4 for all scans.
    \item Correct for the derived residual D-terms by inverting the Jones matrices in Equation~\ref{eq:resmodel}.
    \item Iterate the above steps as many times as specified by users. Utilize the results obtained in Step 6 to generate Stokes $Q$ and $U$ images during Step 1, which are subsequently employed in the remaining calibration steps. The method yields the reduced $\chi^2$ of the fit at every iteration, which can serve as a valuable tool for selecting the appropriate number of iterations.
\end{enumerate}

While this procedure is similar to the instrumental polarization self-calibration procedure implemented in the original GPCAL pipeline \citep{Park2021a}, it differs in that fitting is performed only for leakages of stations that exhibit large residuals between the model and cross-hand visibilities. Using synthetic data sets, we will demonstrate the effectiveness of this strategy in correcting for time-dependent leakages of large amplitudes (Section~\ref{sec:synthetic}). The method is implemented in GPCAL, which is publicly available at \url{https://github.com/jhparkastro/gpcal.}

\section{Validation using Synthetic Data}
\label{sec:synthetic}

To validate the performance of the method, we used synthetic data derived from real VLBA data observed on 2018 Dec 08 at 43 GHz as part of the VLBA-BU-BLAZAR monitoring program\footnote{\url{https://www.bu.edu/blazars/VLBAproject.html}}. Data calibration and analysis methods are described in \pone{}. Stokes $Q$ and $U$ images were produced with CLEAN in Difmap for 10 sources; 3C 279, 3C 273, OJ 287, 3C 454.3, 1633+382, 3C 345, CTA 102, 3C 84, MKN 501, and 0235+164. The sources provide a wide range of total intensity and linear polarization structures and cover a wide range of signal-to-noise ratios.

\begin{figure*}[t!]
\centering
\includegraphics[width = \textwidth]{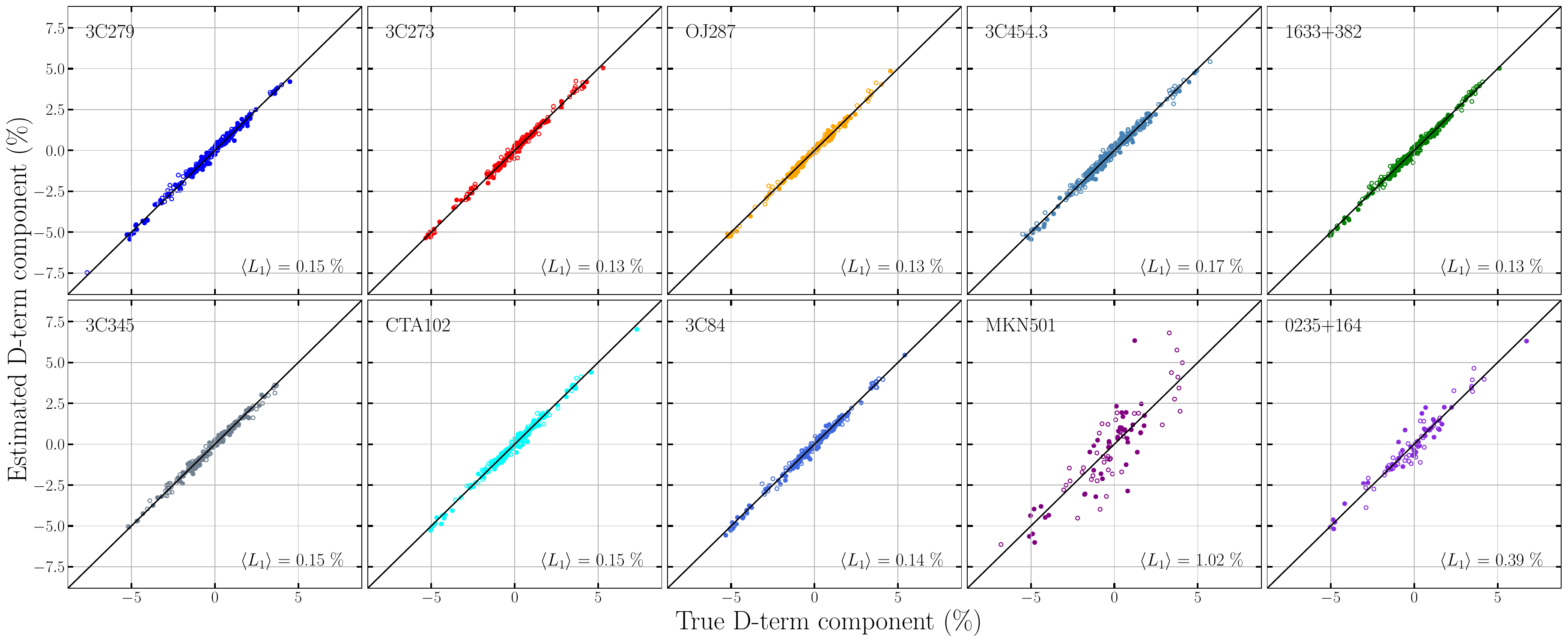}
\caption{Comparison of the time-dependent, ground-truth D-term components (real and imaginary parts) assumed in the synthetic data sets shown on x-axis with the reconstructed D-term components derived by GPCAL shown on y-axis for each scan. Each station's result is presented using distinct colors. For each source, a correlation is shown between the true and estimated D-terms in percentage units. Over all antennas, the $L_1 \equiv |D_{\rm Truth} - D_{\rm Recon}|$ norm is averaged over the real and imaginary components of the D-terms for RCP and LCP. \label{fig:synthetic1to1}}
\end{figure*}

\begin{figure*}[t!]
\centering
\includegraphics[width = \textwidth]{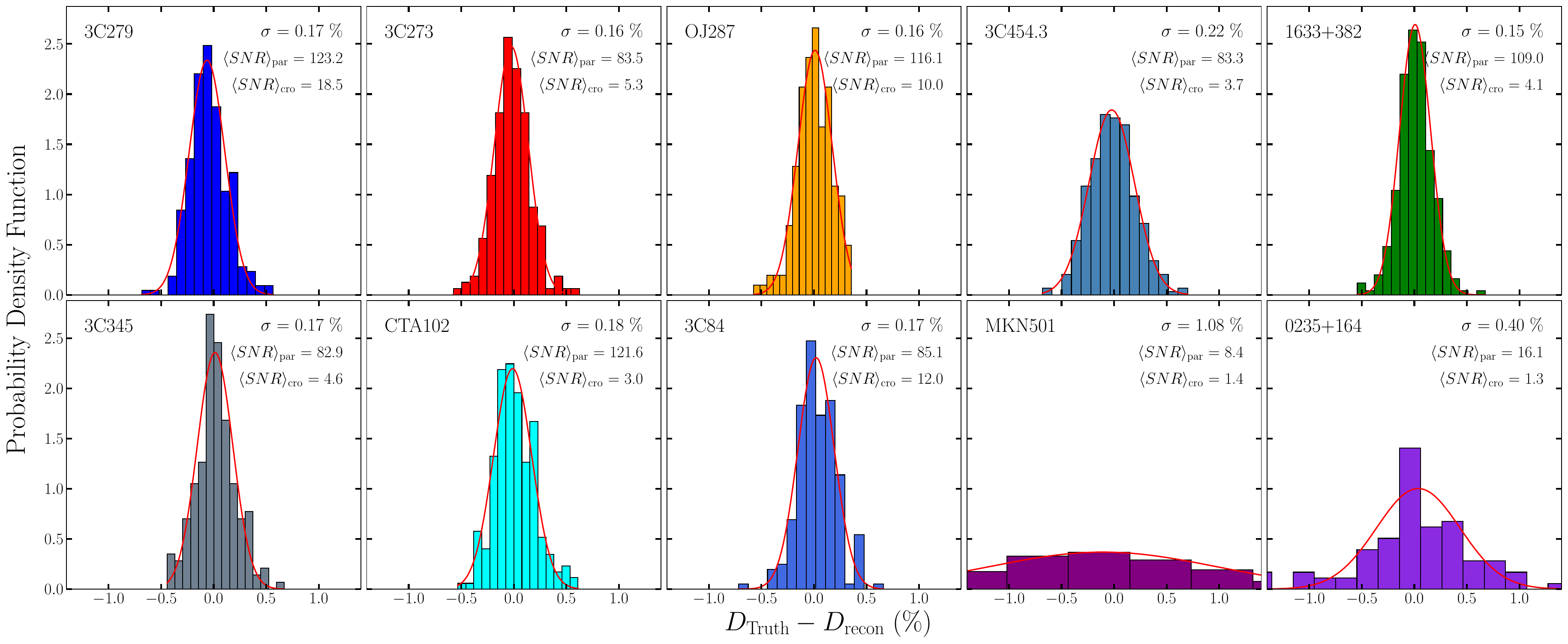}
\caption{PDF of $D_{\rm Truth} - D_{\rm Recon}$ for each source shown in each panel. The red solid lines represent Gaussian distributions fitted to the PDFs. In each panel, the standard deviation ($\sigma$) of the best-fit Gaussian distribution is indicated in percentage units. The average S/N of the parallel-hand ($\langle SNR \rangle_{\rm par}$) and cross-hand visibilities ($\langle SNR \rangle_{\rm cro}$) for each source is denoted in each panel. \label{fig:synthetichisto}}
\end{figure*}

\begin{deluxetable*}{cccccccccccc}
\tablecaption{Goodness of reconstruction of time-dependent leakages by changing the \texttt{timecal\_freepar} parameter\label{tab:freepar}}
\tablewidth{0pt}
\tablehead{\colhead{\texttt{timecal\_freepar}} & \colhead{3C 279} & \colhead{3C 273} & \colhead{OJ 287} & \colhead{3C 454.3} & \colhead{1633+382} & \colhead{3C 345} & \colhead{CTA 102} & \colhead{3C 84} & \colhead{MKN 501} & \colhead{0235+164}
}
\startdata
\hline
1 & 0.17 & 0.17 & 0.17 & 0.19 & 0.14 & 0.19 & 0.19 & 0.15 & 1.28 & 0.30 \\
2 & 0.17 & 0.16 & 0.16 & 0.22 & 0.15 & 0.17 & 0.18 & 0.17 & 1.08 & 0.40 \\
5 & 0.21 & 0.18 & 0.16 & 0.22 & 0.15 & 0.17 & 0.22 & 0.22 & 1.14 & 0.65 \\
10 & 0.50 & 0.76 & 2.16 & 0.84 & 0.81 & 1.64 & 1.56 & 0.73 & 6.37 & 7.27 \\
\hline
\enddata
\tablecomments{Standard deviations of Gaussian distributions fitted to PDFs of $D_{\rm Truth} - D_{\rm Recon}$ are shown in units of \% depending on the \texttt{timecal\_freepar} parameter. Different columns are used to display the results from different sources.}
\end{deluxetable*}

The synthetic data for these 10 sources were generated using GPCAL as described in \cite{Park2021c}. GPCAL generates synthetic data sets using Equation~\ref{eq:matmodel} and the $I$, $Q$, and $U$ CLEAN models as ground truth source structures (i.e., $\tilde{I}$, $\tilde{Q}$, $\tilde{U}$ in Equation~\ref{eq:stokes}), assuming $\tilde{V}=0$. Based on the uncertainties of each real data point, we added thermal noise to the corresponding synthetic data point. Afterwards, the data was averaged over the entire frequency band in order to increase the signal-to-noise ratio. We assumed unity antenna gains, i.e., $\G = \I$. We introduced on-axis D-terms, which are randomly chosen based on the D-term distribution estimated by GPCAL from the real data, and are assumed to remain constant over time. We then introduced off-axis D-terms that vary randomly from scan to scan and whose real and imaginary components follow a Gaussian distribution with a zero mean and a certain standard deviation. As explained in Section~\ref{sec:pipeline} and will be shown in Section~\ref{sec:application}, the off-axis D-terms for most stations are anticipated to have small amplitudes in most cases. However, some stations that experience large pointing errors may have large amplitudes in their off-axis D-terms. In order to reflect this realistic situation in the synthetic data set, we assumed a standard deviation of 0.02 and 0.01 for the VLBA PT and NL station's leakages, respectively, and a standard deviation of 0.002 for the other station's leakages\footnote{Obtaining the precise standard deviation values from actual data is a non-trivial task due to the dependence of off-axis D-term amplitudes on sources and scans. To address this issue, we rely on assuming values that would result in off-axis D-term amplitudes that are reasonably comparable to those obtained from the actual data.}. The method aims to identify and remove off-axis D-terms with large amplitudes from the data.

First, we ran the GPCAL pipeline on the synthetic data set to correct for the on-axis D-terms. Based on the assumption that the source 3C 84 is unpolarized, the pipeline estimates the initial D-terms using the weakly polarized source 3C 84 (with the degree of linear polarization $\lesssim1\%$ at 43 GHz; \citealt{Kim2019}). As a next step, it implements 10 iterations of instrumental polarization self-calibration using bright calibrators with moderate or high linear polarization: 3C 279, 3C 273, 3C 454.3, and OJ 287. The data for 3C 84 is included as well in this procedure under the assumption that it is unpolarized. For all sources, the derived on-axis D-terms are removed from the data. 

Using the method, we corrected for residual time-dependent polarimetric leakages for all simulated sources. We used \texttt{timecal\_freepar} = 2. This means that during each iteration of calibration, the D-terms of the two stations with the largest $\chi^2_m$ values for each scan are free parameters, while the D-terms of the other stations are fixed to be zero (see Steps 2 and 3 in Section~\ref{sec:pipeline}). The time-dependent leakage calibration procedure was repeated ten times. 

We present in Figure~\ref{fig:synthetic1to1} a comparison of the real and imaginary components of the time-dependent D-terms assumed in the synthetic data generation (x-axis) with the reconstructed D-terms estimated by GPCAL (y-axis). The reconstructed D-terms are the sum of the on-axis and off-axis terms. Each panel indicates the average $L_1 \equiv |D_{\rm Truth} - D_{\rm Recon}|$ norm. Our method can reconstruct the true D-terms accurately, except for MKN 501 and 0235+164, which have a low signal-to-noise ratio.

Figure~\ref{fig:synthetichisto} shows the probability density function (PDF) of the difference between the reconstructed and truth D-term components for each source. As indicated in each panel, we fitted a Gaussian function to each PDF and derived its standard deviation ($\sigma$). For most sources, $\sigma$ is close to 0.2\%, which is the standard deviation of a Gaussian distribution used to generate random time-dependent D-term components, except for NL and PT. Consequently, the method can correct for time-dependent D-terms with large amplitudes from some stations, but cannot accurately constrain time-dependent D-terms with small amplitudes. A linear polarization image's quality is primarily limited by the former D-terms, while the latter is naturally expected in most realistic scenarios. Thus, the synthetic data test validates the effectiveness of the method, which is primarily designed to correct for time-dependent leakages with large amplitudes that usually appear at certain stations and/or scans.

For the weak sources MKN 501 and 0235+164, $\sigma$ was substantially greater than 0.2\%. We denote the average S/N of the parallel-hand and cross-hand visibilities for each source in each panel of Figure~\ref{fig:synthetichisto}. It is evident that the S/Ns of the cross-hand visibilities for these sources are notably low\footnote{However, we note that the S/Ns of the parallel-hand visibilities are also crucial for the precise reconstruction of time-dependent leakages. In our method, we assume that $\mathscr{RR}$ and $\mathscr{LL}$ are equivalent to the self-calibrated visibilities $r^{RR}_{\rm cal}$ and $r^{LL}_{\rm cal}$, respectively, as expressed in Equation~\ref{eq:fitmodel}. This is the underlying reason why the reconstruction for 0235+164 exhibits superior results compared to that of MKN 501, despite their cross-hand SNRs being similar to each other.} This result demonstrates that we cannot derive accurate solutions due to the limited signal to noise ratio for these sources despite the small number of parameters available for each round of fitting. Thermal noise would likely limit the quality of linear polarization images of these sources, and systematic errors such as time-dependent leakages would not have a significant impact. As a result, it is not recommended to perform time-dependent leakage calibration on weak sources.

Probably the most critical parameter for the method would be \texttt{timecal\_freepar}, since high values of this parameter would result in large correlations between the fitting parameters and overfitting. As a demonstration of this effect, we applied the method by changing this parameter and presented the sigma values for each source from each run in Table 1. When \texttt{timecal\_freepar} = 2, the sigma values are the lowest for most of the bright sources. For weak sources MKN 501 and 0235+164, the values tend to become smaller as \texttt{timecal\_freepar} decreases. With \texttt{timecal\_freepar} = 10, i.e., when the D-terms of all 10 stations are free parameters, the values become very high for all sources. Using small \texttt{timecal\_freepar} values is recommended for VLBA data at 43 GHz, though the optimal parameter may differ depending on the quality and array of the data.

\section{Application to Real Data}
\label{sec:application}
\subsection{Data \& Analysis}

Using two real VLBA data sets, we evaluate the effectiveness of the method. One is the same data analyzed in \pone{} and in Section~\ref{sec:synthetic} (project code: BM462M) observed on 2018 Dec 08. In the period between August 2018 and October 2019, the VLBA pointing model was not optimized due to problems that occurred during the transition of the telescope control computers \citep{Blanchard2021}. Due to this, the quality of the VLBA data observed during this period was affected, particularly at high frequencies, where the antenna beam size is small. The inaccurate antenna pointing model also affected the BM462M data we analyzed. As a consequence, off-axis D-terms will be pronounced and will significantly limit the quality of linear polarization images derived from this data.

A second set of data is from the M87 monitoring program observed using the VLBA simultaneously at 4.7, 5.1, 6.5, and 6.8 GHz in order to investigate the kinematics of the M87 jet (\citealt{Park2019b}; Park et al. in prep.). We have selected the data observed on 2021 July 02 at 4.7 GHz (project code: BP249H). The VLBA Los Alamos (LA) station reported rain near the end of the observation, and the Stokes $Q$ and $U$ CLEAN models of all sources were not able to fit the data well after the rain began. Weather conditions may affect the leakage characteristics of this station, which motivates us to test our method.

Data postcorrelation was performed with AIPS and hybrid imaging with CLEAN and self-calibration in Difmap. On-axis D-terms as well as frequency-dependent leakages were corrected using GPCAL, as described in \pone{}. Following this, we corrected for time-dependent leakages based on our method, using the parameter \texttt{timecal\_freepar} = 2 and iterating the calibration procedure 10 times.

\begin{figure*}[t!]
\centering
\includegraphics[width = \textwidth]{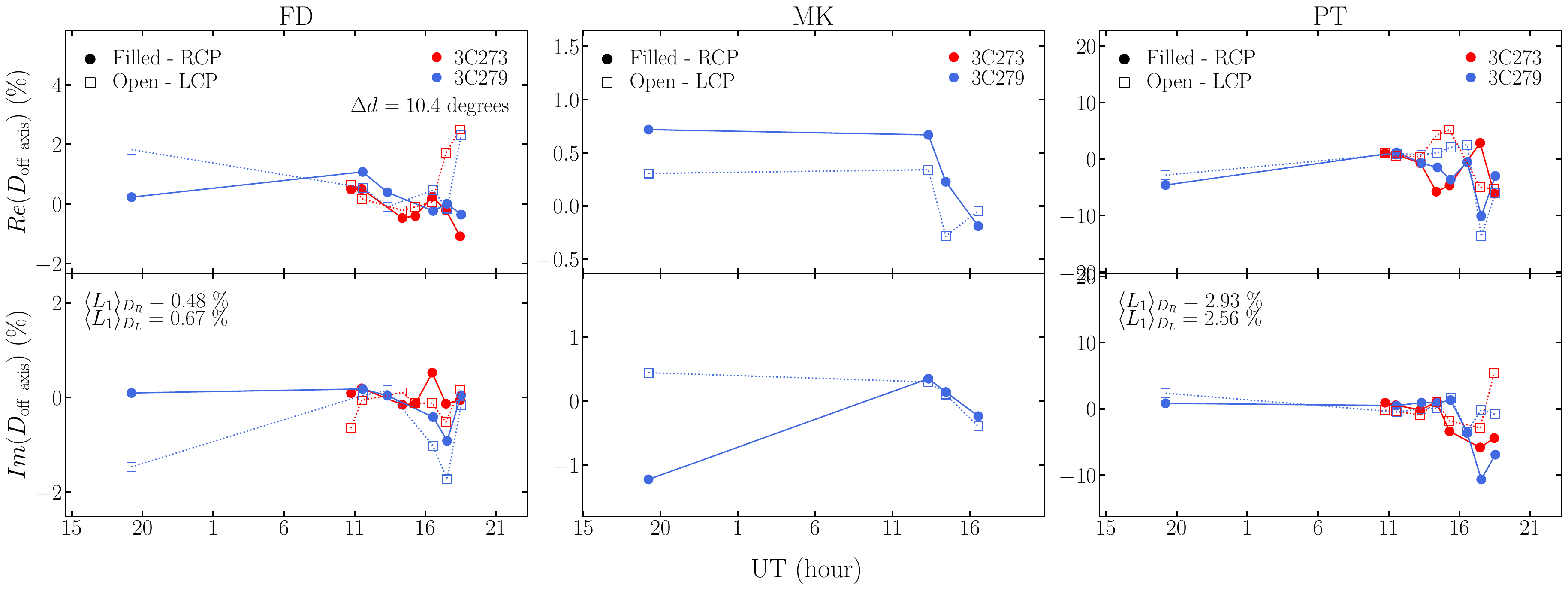}
\includegraphics[width = \textwidth]{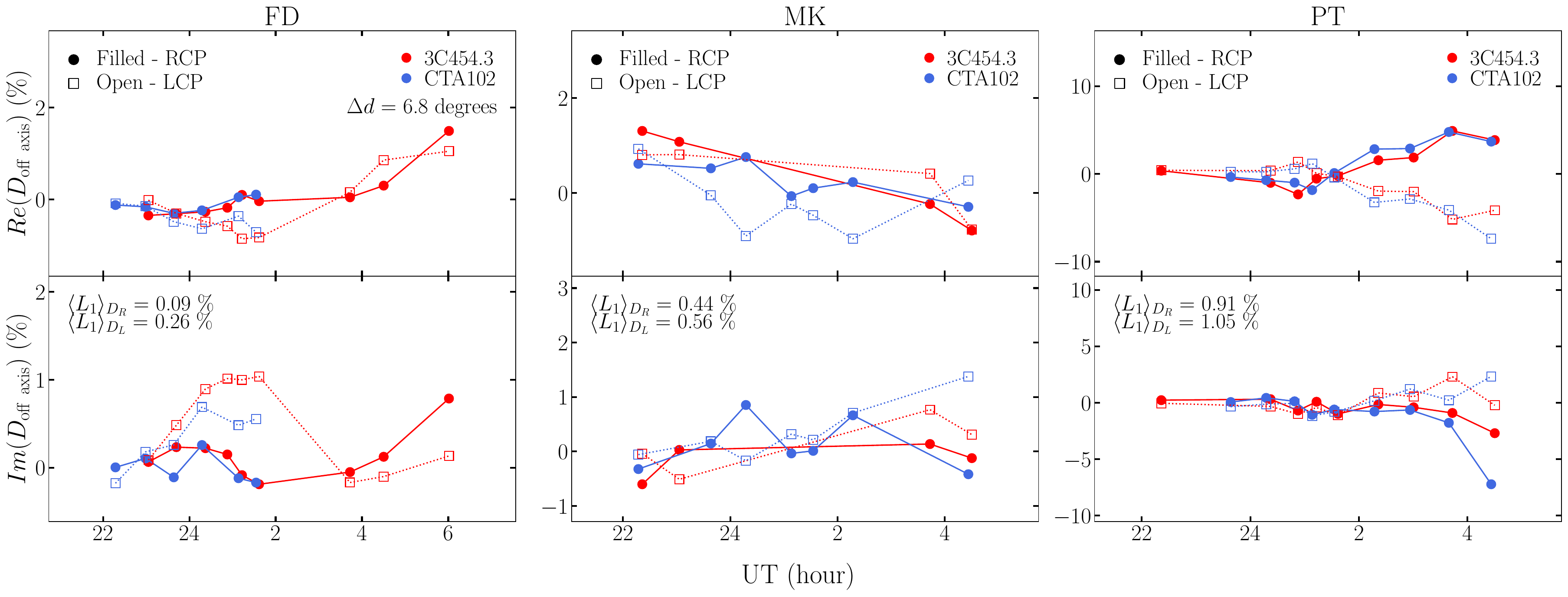}
\includegraphics[width = \textwidth]{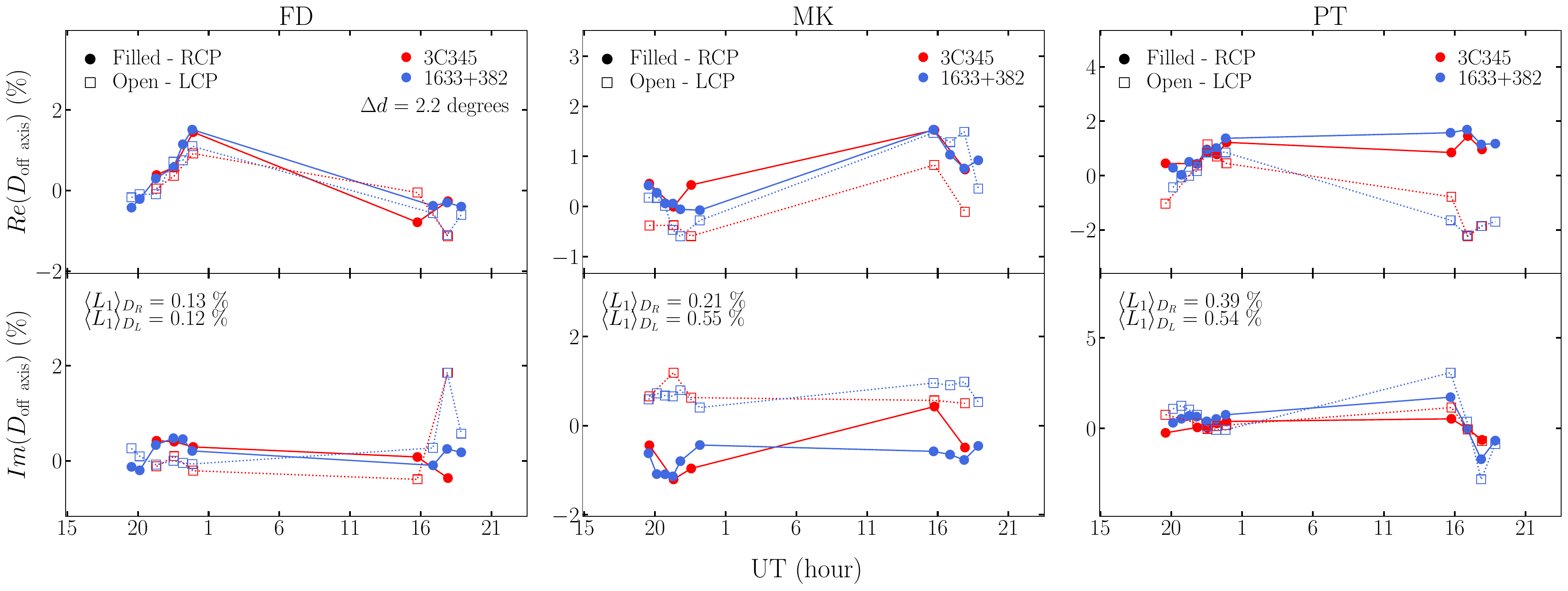}
\caption{The derived off-axis D-terms for the BM462M data as a function of time. In the diagram, the filled circles and open squares represent the D-terms for RCP and LCP, respectively. The upper and lower panels in each panel display the real and imaginary parts of the results for each antenna. Different rows of results are presented for different pairs of adjacent sources in the sky (3C 273 and 3C 279 in the top, 3C 454.3 and CTA 102 in the middle, and 3C 345 and 1633+382 in the bottom). In each figure, the averaged norm $L_1 \equiv |D_i - D_j|$ between the neighboring scans (separated by less than 10 minutes) for the adjacent sources $i$ and $j$ is shown for each polarization. The symbol $\Delta d$ denotes the distance between adjacent sources in the celestial sphere. While we provide results for only three antennas as an illustration, the findings for other antennas were also qualitatively comparable. \label{fig:bm462moffdterm}}
\end{figure*}

\begin{figure*}[t!]
\centering
\includegraphics[width = \textwidth]{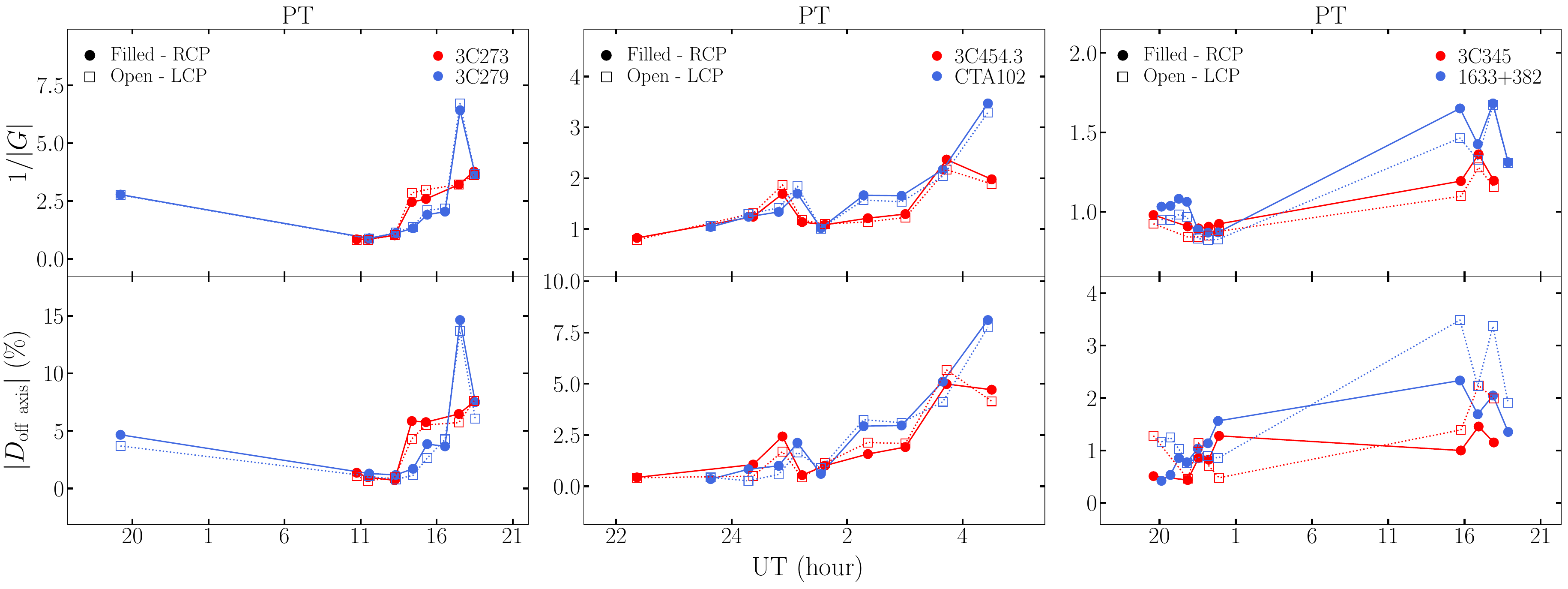}
\caption{Amplitudes of gain correction factors (upper) and off-axis D-terms for the VLBA Pie Town station for different pairs of adjacent sources in the BM462M data set as a function of time. The filled circles and open squares represent the results for RCP and LCP, respectively. \label{fig:bm462mgain}}
\end{figure*}

\begin{figure*}[t!]
\centering
\includegraphics[width = 0.7\textwidth]{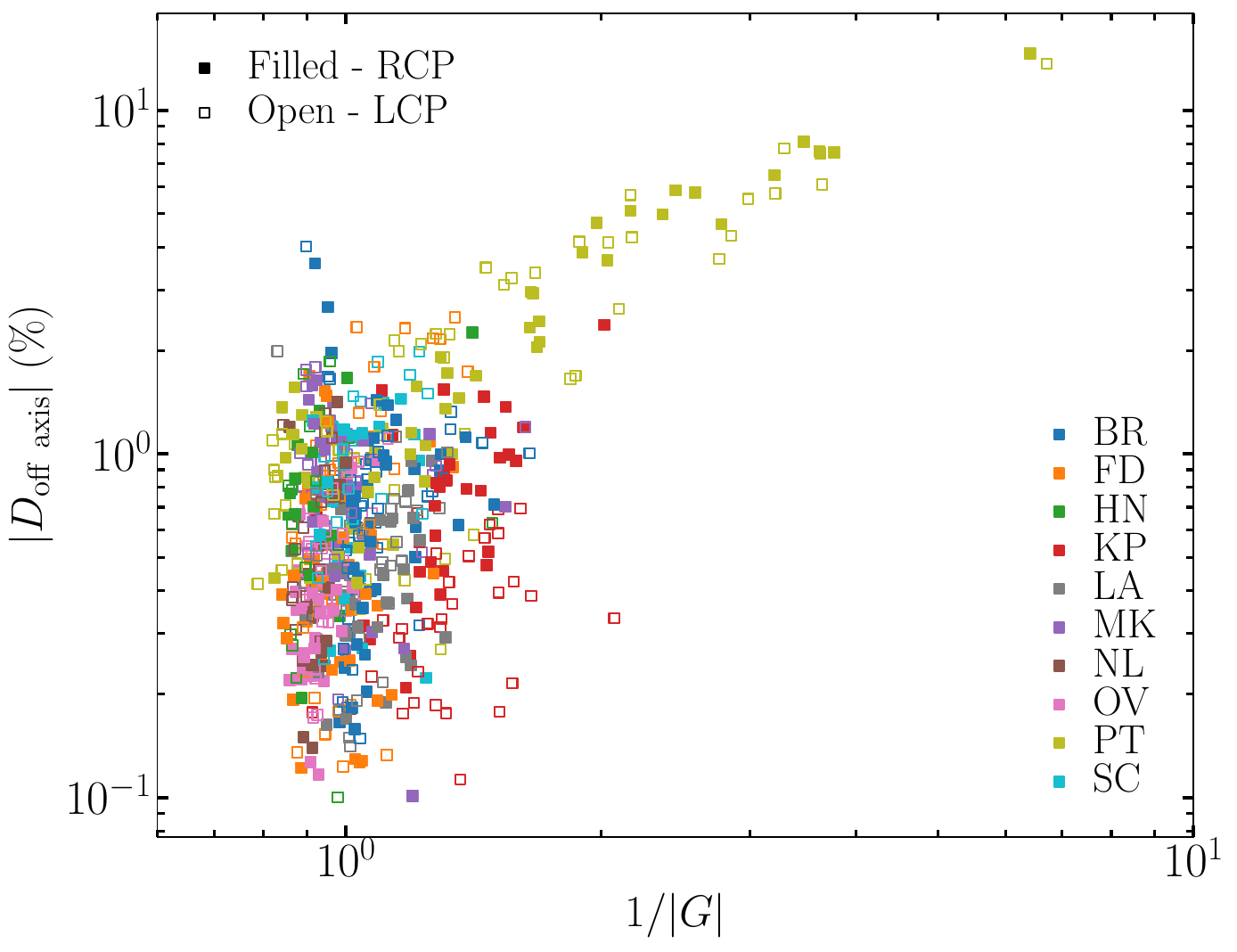}
\caption{Off-axis D-term amplitudes as a function of gain correction factors for the BM462M data set. RCP and LCP are represented by filled circles and open squares, respectively. Different stations are represented by different colors. \label{fig:bm462mgaindtermall}}
\end{figure*}

\subsection{Results}

Figure~\ref{fig:bm462moffdterm} shows the off-axis leakages as a function of time for the BM462M data. Although we present the results for three antennas only as an example, the results for other antennas were also qualitatively similar. The primary purpose of this test is to compare the trend of the off-axis D-terms derived from sources located near each other in the sky. The accuracy of the antenna pointing model depends on the direction of the antenna and the time of day. As a result, nearby scans of nearby sources may also experience similar levels of antenna pointing errors. We can expect to observe similar trends in off-axis D-terms for these scans if the direction-dependent leakages and inaccurate antenna pointing model are the primary causes of time-dependent leakages. We compare the derived off-axis D-terms between 3C 273 and 3C 279 (separated by 10.4 degrees in the sky), 3C 454.3 and CTA 102 (6.8 degrees), and 3C 345 and 1633+382 (2.2 degrees). 

For all antennas, the trends of the derived off-axis D-terms from nearby sources are very similar. In each panel, we present the average $L_1$ norm of the D-term components between nearby scans (within ten minutes) of these source pairs. We obtained smaller average $L_1$ norms for closer pairs of sources, which is reasonable since antenna pointing offsets, and therefore off-axis D-terms, are direction-dependent.

\begin{figure*}[t!]
\centering
\includegraphics[width = \textwidth]{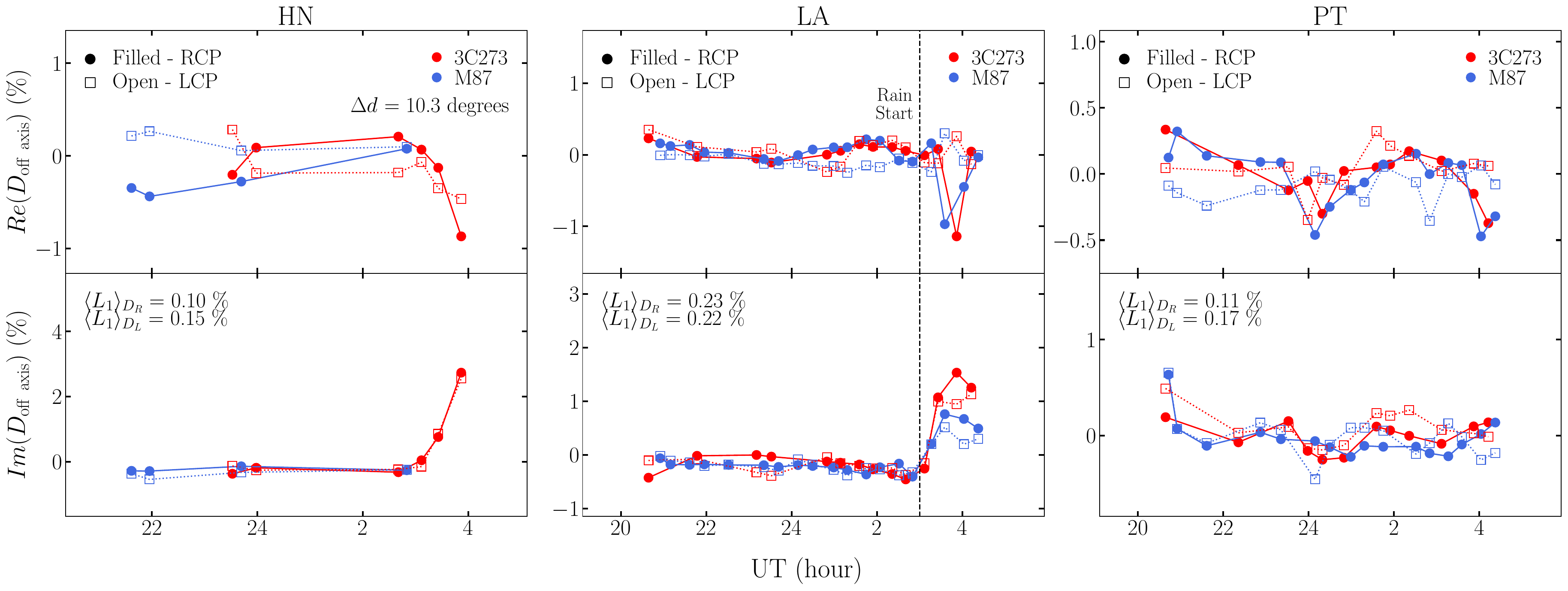}
\caption{Same as Figure~\ref{fig:bm462moffdterm} but for the BP249H data. After UT 3, rain was reported at the VLBA Los Alamos station (LA; middle panel), following which the off-axis D-term amplitudes increased substantially. \label{fig:bp249hoffdterm}}
\end{figure*}

The method did not attempt to derive time-dependent D-terms for the Mauna Kea (MK) station for 3C 273 (the top middle panel in Figure~\ref{fig:bm462moffdterm}). It is due to the fact that the $\chi^2_m$ value for this station is always low during any calibration round. The 3C 273 jet displays a complex linear polarization structure with a moderate to high level of linear polarization \citep[e.g.,][]{Attridge2005, Hada2016, Park2021a}. Thus, the cross-hand visibility of the long baselines associated with MK station has low amplitudes and low signal-to-noise ratios (at a $\lesssim0.1$ Jy level for this data). The effects of systematic errors, such as time-dependent leakages, are not significant for these baselines, and therefore GPCAL does not attempt to model them.

Figure~\ref{fig:bm462mgain} illustrates the amplitudes of the gain-correction factors (1/$|G|$) and the amplitudes of the derived off-axis D-terms as a function of time for the VLBA Pie Town station. As a result of the inaccurate antenna pointing model, this station exhibits large gain correction factors and off-axis D-terms. Both quantities for nearby source pairs exhibit very similar trends as expected. In addition, the trends between the gain correction factors and the off-axis D-terms are also very similar. Due to the direction-dependent nature of antenna leakage, the effects of off-axis D-terms become more prominent for large antenna pointing offsets.

Figure~\ref{fig:bm462mgaindtermall} shows the amplitudes of the derived off-axis D-terms as a function of the gain correction factors for all scans and all sources. Regardless of the antenna, there is a strong positive correlation between the two quantities. It is demonstrated that time-dependent leakages are present in the data due to the direction-dependent D-terms and imperfect pointing offsets. Their amplitudes can be very large for scans that are affected by large pointing offsets.

In Figure~\ref{fig:bp249hoffdterm}, we present the derived off-axis D-terms as a function of time for 3C 273 and M87, separated by 10.3 degrees in the sky, for the BP249H data. In line with the results obtained for BM462M data, the off-axis D-terms for these sources also exhibit similar trends. As the antenna pointing is accurate at this low frequency (4.7 GHz), the amplitudes of the off-axis D-terms are generally small. For LA station, however, large off-axis D-terms are observed after around 3 UT, when rainfall began. Both 3C 273 and M87 exhibited this trend, indicating that severe weather had an impact on the large off-axis D-terms. Observations at other frequencies have also yielded similar results, although they are not included in the present paper.

\begin{figure*}[t!]
\centering
\includegraphics[width = \textwidth]{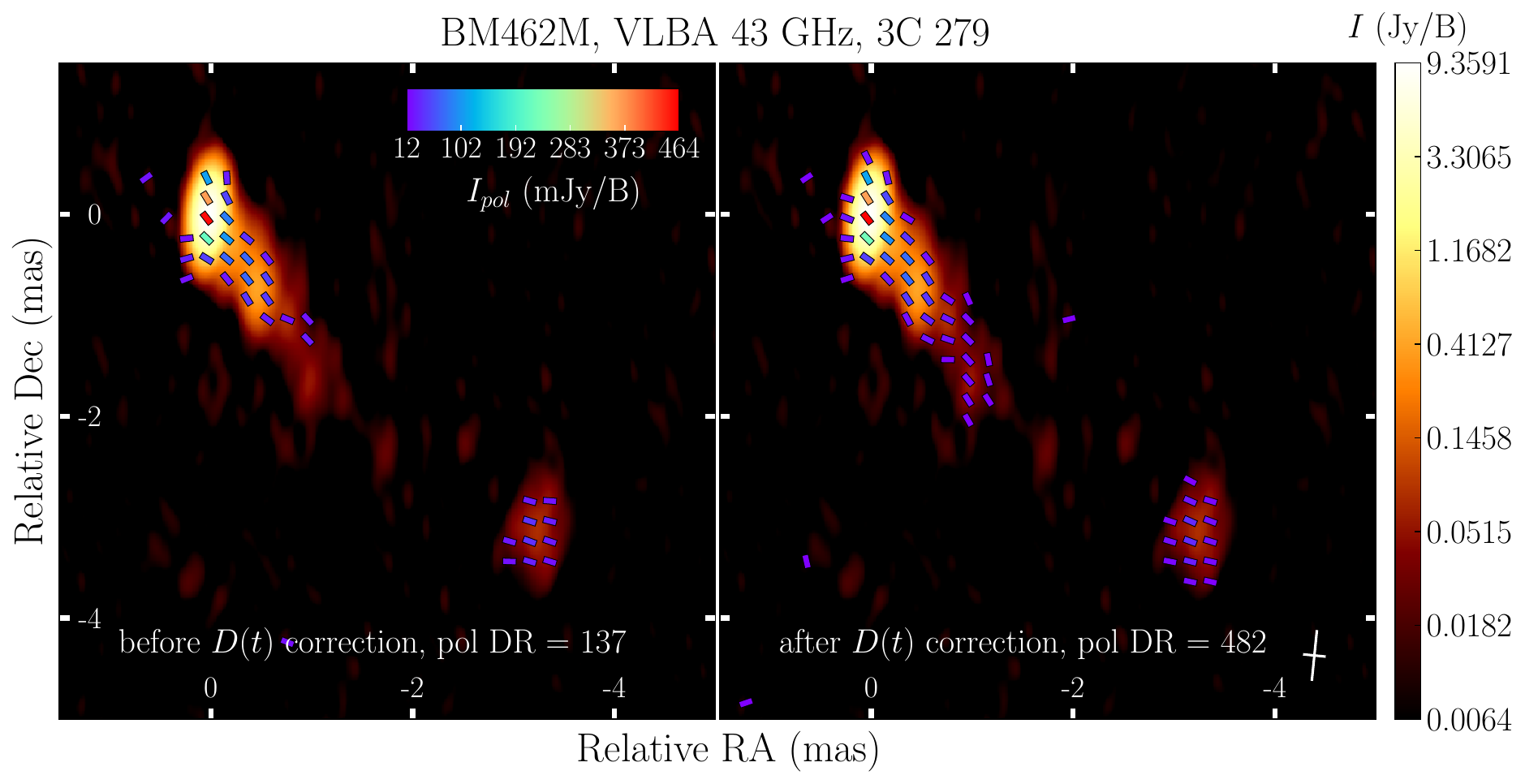}
\includegraphics[trim = 0mm 5mm 0mm 0mm, width = \textwidth]{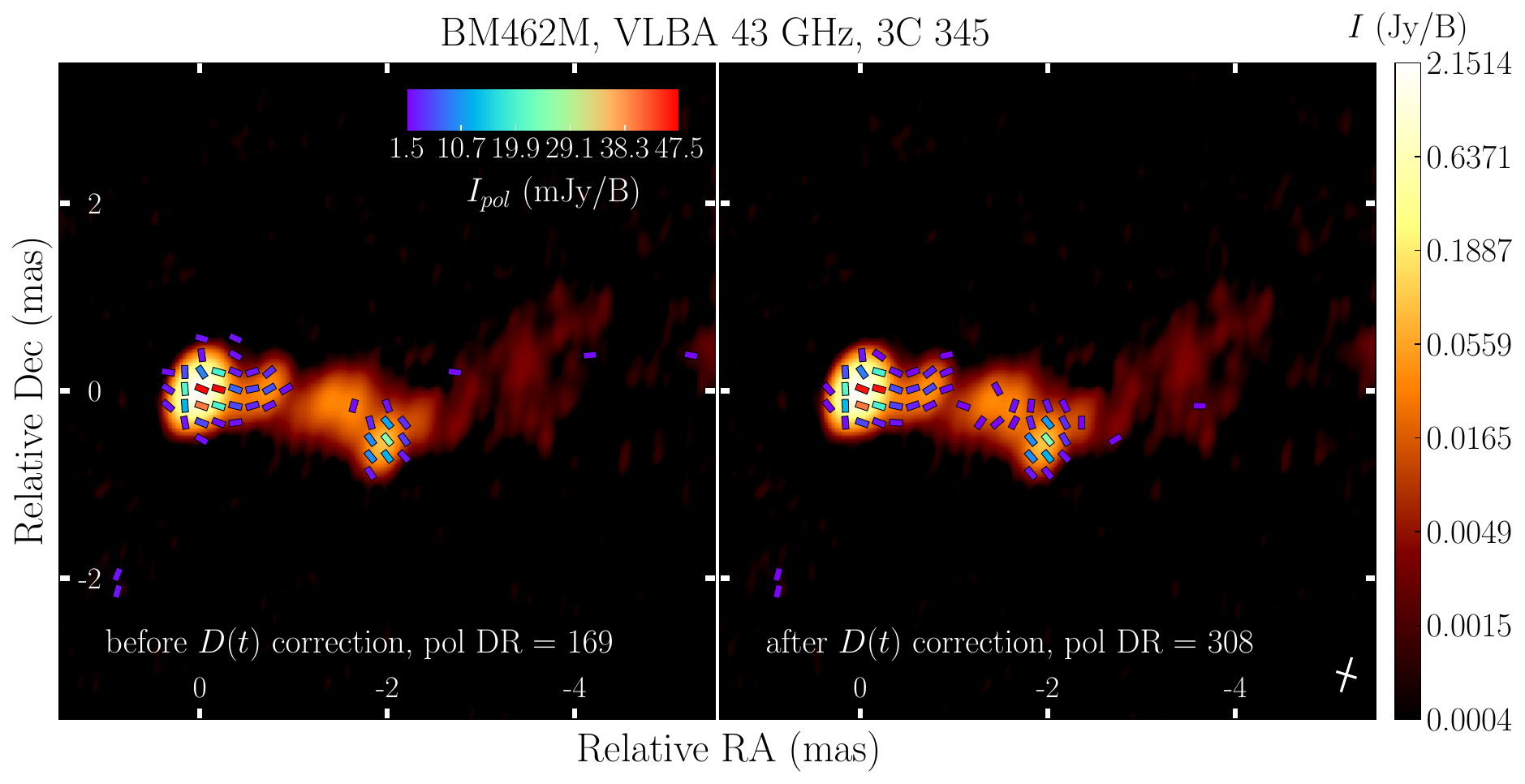}
\caption{Linear polarization images of 3C 279 (upper) and 3C 345 (lower) from the BM462M data set. The hot color indicates the total intensity distribution and the colored ticks indicate the electric vector position angles (EVPAs), with the color scales indicating Ricean de-biased linearly polarized intensity. Presented in the left and right panels are images obtained before and after time-dependent leakage correction. There is a note in each panel indicating the linear polarization dynamic range, which is defined as the ratio of the peak linear polarization intensity to the linear polarization off-source rms noise. In the bottom right corner, the shape of the synthesized beam is shown.\label{fig:bm462mmap}}
\end{figure*}

\begin{figure*}[t!]
\centering
\includegraphics[width = \textwidth]{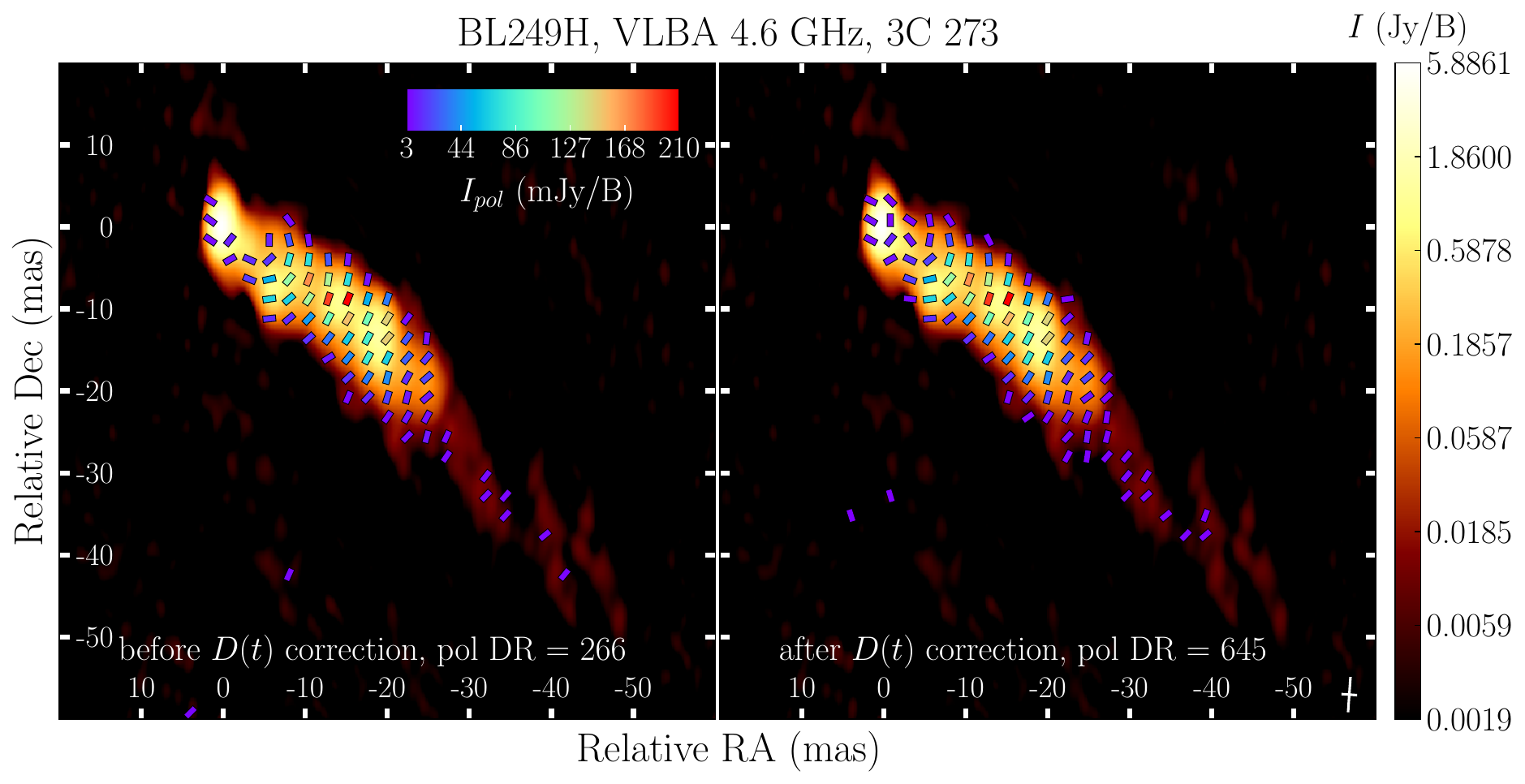}
\includegraphics[width = \textwidth]{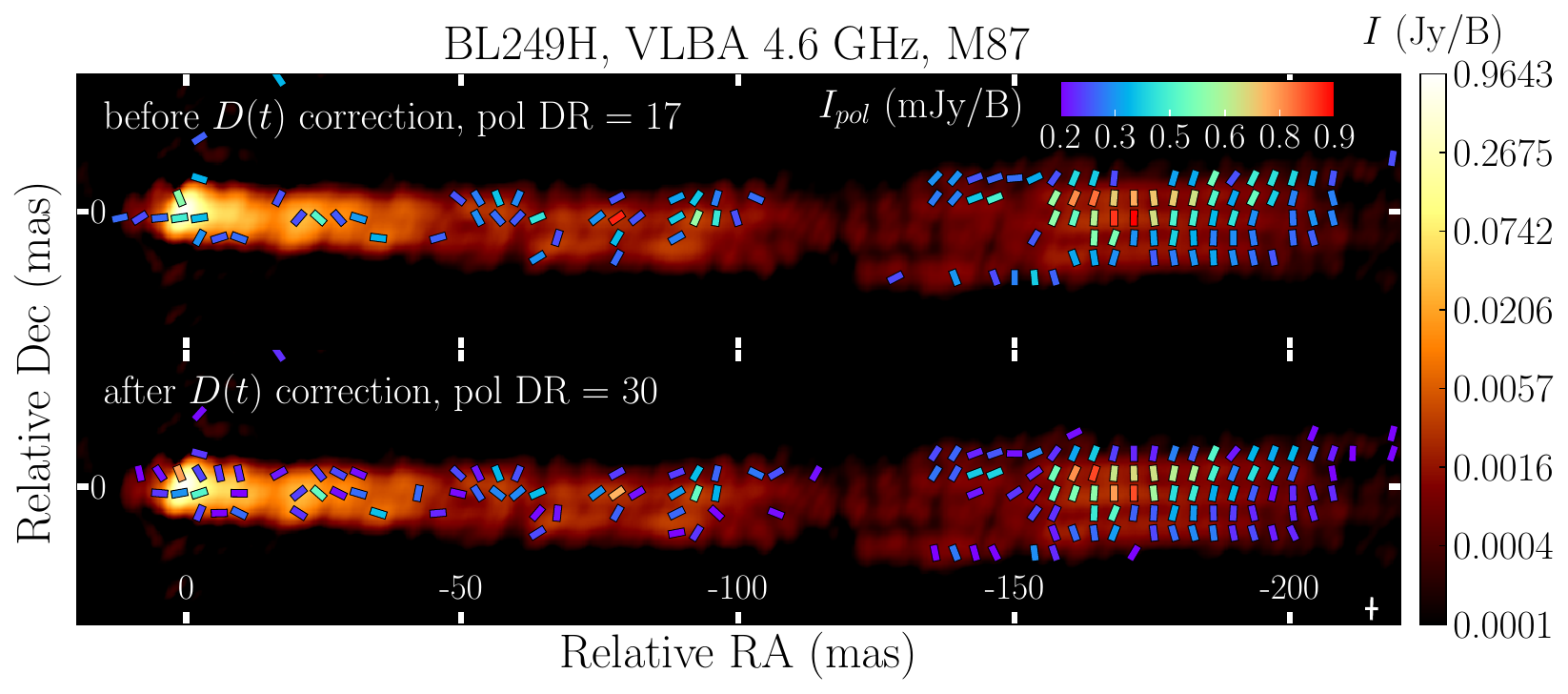}
\caption{Same as Figure~\ref{fig:bm462mmap} but for the BP249H data for 3C 273 (upper) and M87 (lower). \label{fig:bp249hmap}}
\end{figure*}

\subsection{Evaluation}
\label{sec:evaluation}

Figure~\ref{fig:bm462mmap} illustrates the linear polarization images of 3C 279 and 3C 345 derived from the BM462M data before (left) and after (right) time-dependent leakage correction. A Ricean de-biasing correction has been applied to the images (i.e., $P_{\rm corr} = P_{\rm obs}\sqrt{1 - (\sigma_P / P_{\rm obs})}$), where $P_{\rm corr}$ and $P_{\rm obs}$ denote the corrected and original linearly polarized intensities \citep{WK1974, Lister2018}, respectively, and $\sigma_P \equiv (\sigma_{Q} + \sigma_{U}) / 2$ is the average of the rms noise in the off-source regions in Stokes $Q$ and $U$ images \citep{Hovatta2012}. In each image, we note the polarization dynamic range, which is defined as the ratio of the peak linear polarization intensity to $\sigma_P$. For each image, linear polarization vectors are drawn for linearly polarized intensities exceeding three times the $\sigma_P$ value.

Through the use of the time-dependent leakage correction method, the noise level of the polarization images has been significantly reduced, resulting in an increase in polarization dynamic range by factors of two to three. After the time-dependent leakage correction, linear polarization emission was observed in the jet regions of weak total intensity that was not evident in the original linear polarization images. In the case of 3C 279, which has a high linearly polarized flux ($\sim0.8$ Jy), the noise level in the polarization images is more dominated by systematic errors than by thermal noise. Using the method, residual systematic errors in the data are removed, making it possible to detect the weak polarization emission in the jet.

A comparison of linear polarization images of 3C 273 and M87 based on the BP249H data with and without time-dependent leakage correction is presented in Figure~\ref{fig:bp249hmap}. Because of the improvement in polarization dynamic range, weaker linear polarization emission was detected in both sources, similar to the results obtained from the BM462M data.

\section{Conclusions}
\label{sec:conclusion}

In this series of papers, we present new methods for calibrating frequency and time-dependent leakages in VLBI data, which have been implemented in GPCAL. A wide fractional bandwidth is provided by modern VLBI arrays. The instrumental polarization is affected by chromatic effects since telescope systems are usually designed to produce the most optimal polarization response at a given nominal frequency \citep[e.g.,][]{MartiVidal2021}. In \pone{}, we present a method for correcting leakages that vary in frequency. It is also possible for instrumental polarization to change over time. The reason for this is that instrumental polarization is a direction-dependent effect \citep[e.g.,][]{Smirnov2011}. Therefore, instrumental polarization can vary depending on the accuracy of the antenna pointing. Antenna pointing is subject to uncertainty, and can be adversely affected by the weather, such as strong winds and the deformation of antennas caused by sunlight, which may result in instrumental polarization that is time-dependent. In poor weather conditions and with large dishes, this effect is more pronounced. The purpose of this paper is to introduce a method for correcting time-dependent leakages.

This method works on the data after correcting the "on-axis" D-terms, which are calculated assuming that the D-terms are constant during observation. In this manner, "residual" leakages are derived. It calculates how well the Stokes $Q$ and $U$ images fit the cross-hand visibilities for each antenna and for each scan based on the data. Next, it determines which antennas exhibit the greatest difference between the data and the model. By assuming that the D-terms of all other antennas are zero, the method determines the best-fit D-terms for these antennas\footnote{We plan to implement another calibration strategy that will correct for off-axis leakages only for scans and stations that are affected by large antenna pointing offsets in the future. It is based on the tight correlation between the antenna gain correction factors and the amplitudes of the derived off-axis D-terms (Figures~\ref{fig:bm462mgain} and ~\ref{fig:bm462mgaindtermall}). This strategy may reduce the degrees of freedom in the fitting process, but it requires good images of the source's total intensity.}. The procedure is iterated until the solutions are converged, and the number of iterations can be controlled by the user.

The method was tested using a synthetic data set containing time-dependent leakages generated from real VLBA data observed at 43 GHz. Leakages consist of two components: the on-axis, stable D-terms drawn from the distribution of the D-terms from the real data, as well as the off-axis, variable D-terms that are randomly selected from Gaussian distributions for each scan. It is assumed that two antennas have large standard deviations (0.02 and 0.01 for PT and NL stations, respectively) for the Gaussian distributions and other antennas have small standard deviations (0.002), in order to mimic some recently observed VLBA data sets at 43 GHz. 

Based on the method, we derived time-dependent leakages for the synthetic data set. We found that the method was able to successfully reconstruct the time-dependent leakages that were assumed in the generation of synthetic data. In most cases, the distribution of the difference between the estimated and ground-truth D-term components (the real and imaginary parts) could be described by Gaussian distributions with standard deviations less than 0.002. This result indicates that the method is capable of capturing the time-dependent leakages of large amplitudes, which are the dominant factors in limiting the quality of linear polarization images. The purpose of developing the method is actually to correct for the largest systematic errors in the data by using prior information that the off-axis D-term amplitudes of most antennas in realistic VLBI arrays are usually small. 

Our method was applied to two sets of real data obtained with the VLBA at 4.7 and 43 GHz. In the VLBA 4.7 GHz data, significant deviations are observed between the source polarization models and cross-hand visibilities for all baselines associated with the VLBA Los Alamos station after the time the station reported rain. An antenna control computer upgrade caused inaccurate antenna pointing models in the VLBA 43 GHz data. We derived the off-axis D-terms for several pairs of sources located close to one another in the sky as a function of time. The trends of the derived off-axis D-terms between nearby sources are very similar. Off-axis D-terms for nearby scans of less separated sources displayed a higher degree of consistency. Additionally, the amplitudes of the off-axis D-terms and the gain-correction factors show very similar trends. Based on these results, it can be concluded that the method is capable of capturing time-dependent leakages caused by direction-dependent leakages and imperfect antenna pointing. 

For all sources analyzed in this paper, the dynamic ranges of linear polarization images have been improved by factors of $\gtrsim2$ after correcting for time-dependent leakages using the method. According to these results, the method can significantly enhance the polarization image fidelity of VLBI data. Additionally, the method will be useful for improving global VLBI arrays operating at millimeter wavelengths, such as the GMVA and EHT, which have very small antenna beams because of the high observing frequencies and large dishes.

\acknowledgments

We express our gratitude to the anonymous referee for conducting a comprehensive review of our manuscript, which greatly enhanced the quality of the paper. J.P. acknowledges financial support through the EACOA Fellowship awarded by the East Asia Core Observatories Association, which consists of the Academia Sinica Institute of Astronomy and Astrophysics, the National Astronomical Observatory of Japan, Center for Astronomical Mega-Science, Chinese Academy of Sciences, and the Korea Astronomy and Space Science Institute. This work is supported by the Ministry of Science and Technology of Taiwan grant MOST 109-2112-M-001-025 and 108-2112-M-001-051 (K.A). The VLBA is an instrument of the National Radio Astronomy Observatory. The National Radio Astronomy Observatory is a facility of the National Science Foundation operated by Associated Universities, Inc. 

\facilities{VLBA (NRAO)}

\software{AIPS \citep{Greisen2003}, Difmap \citep{Shepherd1997}, GPCAL \citep{Park2021a}, ParselTongue \citep{Kettenis2006}, Scipy \citep{Scipy2020}}

\clearpage

\bibliography{AAS44257R1}{}

\begin{thebibliography}{}
\expandafter\ifx\csname natexlab\endcsname\relax\def\natexlab#1{#1}\fi
\providecommand{\url}[1]{\href{#1}{#1}}
\providecommand{\dodoi}[1]{doi:~\href{http://doi.org/#1}{\nolinkurl{#1}}}
\providecommand{\doeprint}[1]{\href{http://ascl.net/#1}{\nolinkurl{http://ascl.net/#1}}}
\providecommand{\doarXiv}[1]{\href{https://arxiv.org/abs/#1}{\nolinkurl{https://arxiv.org/abs/#1}}}

\bibitem[{{Asada} {et~al.}(2002){Asada}, {Inoue}, {Uchida}, {Kameno},
  {Fujisawa}, {Iguchi}, \& {Mutoh}}]{Asada2002}
{Asada}, K., {Inoue}, M., {Uchida}, Y., {et~al.} 2002, \pasj, 54, L39,
  \dodoi{10.1093/pasj/54.3.L39}

\bibitem[{{Attridge} {et~al.}(2005){Attridge}, {Wardle}, \&
  {Homan}}]{Attridge2005}
{Attridge}, J.~M., {Wardle}, J. F.~C., \& {Homan}, D.~C. 2005, \apjl, 633, L85,
  \dodoi{10.1086/498392}

\bibitem[{{Blanchard}(2021)}]{Blanchard2021}
{Blanchard}, J. 2021, VLBA Test Memo 73 (Socorro: NRAO)

\bibitem[{{Broderick} {et~al.}(2020){Broderick}, {Gold}, {Karami},
  {Preciado-L{\'o}pez}, {Tiede}, {Pu}, {Akiyama}, {Alberdi}, {Alef}, {Asada},
  {Azulay}, {Baczko}, {Balokovi{\'c}}, {Barrett}, {Bintley}, {Blackburn},
  {Boland}, {Bouman}, {Bower}, {Bremer}, {Brinkerink}, {Brissenden}, {Britzen},
  {Broguiere}, {Bronzwaer}, {Byun}, {Carlstrom}, {Chael}, {Chatterjee},
  {Chatterjee}, {Chen}, {Chen}, {Cho}, {Conway}, {Cordes}, {Crew}, {Cui},
  {Davelaar}, {De Laurentis}, {Deane}, {Dempsey}, {Desvignes}, {Doeleman},
  {Eatough}, {Falcke}, {Fish}, {Fomalont}, {Fraga-Encinas}, {Friberg}, {Fromm},
  {Galison}, {Gammie}, {Garc{\'\i}a}, {Gentaz}, {Georgiev}, {Goddi},
  {G{\'o}mez}, {Gu}, {Gurwell}, {Hada}, {Hecht}, {Hesper}, {Ho}, {Ho}, {Honma},
  {Huang}, {Huang}, {Hughes}, {Inoue}, {Issaoun}, {James}, {Janssen}, {Jeter},
  {Jiang}, {Jim{\'e}nez-Rosales}, {Johnson}, {Jorstad}, {Jung}, {Karuppusamy},
  {Kawashima}, {Keating}, {Kettenis}, {Kim}, {Kim}, {Kino}, {Koay}, {Koch},
  {Koyama}, {Kramer}, {Kramer}, {Krichbaum}, {Kuo}, {Lee}, {Li}, {Li},
  {Lindqvist}, {Lico}, {Liu}, {Liuzzo}, {Lo}, {Lobanov}, {Loinard}, {Lonsdale},
  {Lu}, {MacDonald}, {Mao}, {Marscher}, {Mart{\'\i}-Vidal}, {Matsushita},
  {Matthews}, {Menten}, {Mizuno}, {Mizuno}, {Moran}, {Moriyama},
  {Moscibrodzka}, {M{\"u}ller}, {Nagai}, {Nagar}, {Nakamura}, {Narayan},
  {Narayanan}, {Natarajan}, {Neri}, {Ni}, {Noutsos}, {Okino}, {Olivares},
  {Ortiz-Le{\'o}n}, {Oyama}, {Palumbo}, {Park}, {Pen}, {Pesce}, {Pi{\'e}tu},
  {Plambeck}, {PopStefanija}, {Porth}, {Prather}, {Ramakrishnan}, {Rao},
  {Rawlings}, {Raymond}, {Rezzolla}, {Ripperda}, {Roelofs}, {Rogers}, {Ros},
  {Rose}, {Rottmann}, {Ruszczyk}, {Ryan}, {Rygl}, {S{\'a}nchez},
  {S{\'a}nchez-Arguelles}, {Sasada}, {Savolainen}, {Schloerb}, {Schuster},
  {Shao}, {Shen}, {Small}, {Sohn}, {SooHoo}, {Tazaki}, {Tilanus}, {Titus},
  {Toma}, {Torne}, {Traianou}, {Trippe}, {Tsuda}, {van Bemmel}, {van
  Langevelde}, {van Rossum}, {Wagner}, {Wardle}, {Weintroub}, {Wex}, {Wharton},
  {Wielgus}, {Wong}, {Wu}, {Yoon}, {Young}, {Young}, {Younsi}, {Yuan}, {Yuan},
  {Zensus}, {Zhao}, {Zhao}, {Zhu}, \& {Event Horizon Telescope
  Collaboration}}]{Broderick2020}
{Broderick}, A.~E., {Gold}, R., {Karami}, M., {et~al.} 2020, \apj, 897, 139,
  \dodoi{10.3847/1538-4357/ab91a4}

\bibitem[{{Casadio} {et~al.}(2017){Casadio}, {Krichbaum}, {Marscher},
  {Jorstad}, {G{\'o}mez}, {Agudo}, {Bach}, {Kim}, {Hodgson}, \&
  {Zensus}}]{Casadio2017}
{Casadio}, C., {Krichbaum}, T., {Marscher}, A., {et~al.} 2017, Galaxies, 5, 67,
  \dodoi{10.3390/galaxies5040067}

\bibitem[{{Chael} {et~al.}(2018){Chael}, {Johnson}, {Bouman}, {Blackburn},
  {Akiyama}, \& {Narayan}}]{Chael2018}
{Chael}, A.~A., {Johnson}, M.~D., {Bouman}, K.~L., {et~al.} 2018, \apj, 857,
  23, \dodoi{10.3847/1538-4357/aab6a8}

\bibitem[{{Chael} {et~al.}(2016){Chael}, {Johnson}, {Narayan}, {Doeleman},
  {Wardle}, \& {Bouman}}]{Chael2016}
{Chael}, A.~A., {Johnson}, M.~D., {Narayan}, R., {et~al.} 2016, \apj, 829, 11,
  \dodoi{10.3847/0004-637X/829/1/11}

\bibitem[{{Cotton}(1993)}]{Cotton1993}
{Cotton}, W.~D. 1993, \aj, 106, 1241, \dodoi{10.1086/116723}

\bibitem[{{Event Horizon Telescope Collaboration}
  {et~al.}(2019{\natexlab{a}}){Event Horizon Telescope Collaboration},
  {Akiyama}, {Alberdi}, {Alef}, {Asada}, {Azulay}, {Baczko}, {Ball},
  {Balokovi{\'c}}, {Barrett}, {Bintley}, {Blackburn}, {Boland}, {Bouman},
  {Bower}, {Bremer}, {Brinkerink}, {Brissenden}, {Britzen}, {Broderick},
  {Broguiere}, {Bronzwaer}, {Byun}, {Carlstrom}, {Chael}, {Chan}, {Chatterjee},
  {Chatterjee}, {Chen}, {Chen}, {Cho}, {Christian}, {Conway}, {Cordes}, {Crew},
  {Cui}, {Davelaar}, {De Laurentis}, {Deane}, {Dempsey}, {Desvignes}, {Dexter},
  {Doeleman}, {Eatough}, {Falcke}, {Fish}, {Fomalont}, {Fraga-Encinas},
  {Freeman}, {Friberg}, {Fromm}, {G{\'o}mez}, {Galison}, {Gammie},
  {Garc{\'\i}a}, {Gentaz}, {Georgiev}, {Goddi}, {Gold}, {Gu}, {Gurwell},
  {Hada}, {Hecht}, {Hesper}, {Ho}, {Ho}, {Honma}, {Huang}, {Huang}, {Hughes},
  {Ikeda}, {Inoue}, {Issaoun}, {James}, {Jannuzi}, {Janssen}, {Jeter}, {Jiang},
  {Johnson}, {Jorstad}, {Jung}, {Karami}, {Karuppusamy}, {Kawashima},
  {Keating}, {Kettenis}, {Kim}, {Kim}, {Kim}, {Kino}, {Koay}, {Koch}, {Koyama},
  {Kramer}, {Kramer}, {Krichbaum}, {Kuo}, {Lauer}, {Lee}, {Li}, {Li},
  {Lindqvist}, {Liu}, {Liuzzo}, {Lo}, {Lobanov}, {Loinard}, {Lonsdale}, {Lu},
  {MacDonald}, {Mao}, {Markoff}, {Marrone}, {Marscher}, {Mart{\'\i}-Vidal},
  {Matsushita}, {Matthews}, {Medeiros}, {Menten}, {Mizuno}, {Mizuno}, {Moran},
  {Moriyama}, {Moscibrodzka}, {M{\"u}ller}, {Nagai}, {Nagar}, {Nakamura},
  {Narayan}, {Narayanan}, {Natarajan}, {Neri}, {Ni}, {Noutsos}, {Okino},
  {Olivares}, {Ortiz-Le{\'o}n}, {Oyama}, {{\"O}zel}, {Palumbo}, {Patel}, {Pen},
  {Pesce}, {Pi{\'e}tu}, {Plambeck}, {PopStefanija}, {Porth}, {Prather},
  {Preciado-L{\'o}pez}, {Psaltis}, {Pu}, {Ramakrishnan}, {Rao}, {Rawlings},
  {Raymond}, {Rezzolla}, {Ripperda}, {Roelofs}, {Rogers}, {Ros}, {Rose},
  {Roshanineshat}, {Rottmann}, {Roy}, {Ruszczyk}, {Ryan}, {Rygl},
  {S{\'a}nchez}, {S{\'a}nchez-Arguelles}, {Sasada}, {Savolainen}, {Schloerb},
  {Schuster}, {Shao}, {Shen}, {Small}, {Sohn}, {SooHoo}, {Tazaki}, {Tiede},
  {Tilanus}, {Titus}, {Toma}, {Torne}, {Trent}, {Trippe}, {Tsuda}, {van
  Bemmel}, {van Langevelde}, {van Rossum}, {Wagner}, {Wardle}, {Weintroub},
  {Wex}, {Wharton}, {Wielgus}, {Wong}, {Wu}, {Young}, {Young}, {Younsi},
  {Yuan}, {Yuan}, {Zensus}, {Zhao}, {Zhao}, {Zhu}, {Algaba}, {Allardi},
  {Amestica}, {Anczarski}, {Bach}, {Baganoff}, {Beaudoin}, {Benson},
  {Berthold}, {Blanchard}, {Blundell}, {Bustamente}, {Cappallo},
  {Castillo-Dom{\'\i}nguez}, {Chang}, {Chang}, {Chang}, {Chen}, {Chilson},
  {Chuter}, {C{\'o}rdova Rosado}, {Coulson}, {Crawford}, {Crowley}, {David},
  {Derome}, {Dexter}, {Dornbusch}, {Dudevoir}, {Dzib}, {Eckart}, {Eckert},
  {Erickson}, {Everett}, {Faber}, {Farah}, {Fath}, {Folkers}, {Forbes},
  {Freund}, {G{\'o}mez-Ruiz}, {Gale}, {Gao}, {Geertsema}, {Graham}, {Greer},
  {Grosslein}, {Gueth}, {Haggard}, {Halverson}, {Han}, {Han}, {Hao},
  {Hasegawa}, {Henning}, {Hern{\'a}ndez-G{\'o}mez}, {Herrero-Illana},
  {Heyminck}, {Hirota}, {Hoge}, {Huang}, {Impellizzeri}, {Jiang}, {Kamble},
  {Keisler}, {Kimura}, {Kono}, {Kubo}, {Kuroda}, {Lacasse}, {Laing}, {Leitch},
  {Li}, {Lin}, {Liu}, {Liu}, {Lu}, {Marson}, {Martin-Cocher}, {Massingill},
  {Matulonis}, {McColl}, {McWhirter}, {Messias}, {Meyer-Zhao}, {Michalik},
  {Monta{\~n}a}, {Montgomerie}, {Mora-Klein}, {Muders}, {Nadolski}, {Navarro},
  {Neilsen}, {Nguyen}, {Nishioka}, {Norton}, {Nowak}, {Nystrom}, {Ogawa},
  {Oshiro}, {Oyama}, {Parsons}, {Paine}, {Pe{\~n}alver}, {Phillips}, {Poirier},
  {Pradel}, {Primiani}, {Raffin}, {Rahlin}, {Reiland}, {Risacher}, {Ruiz},
  {S{\'a}ez-Mada{\'\i}n}, {Sassella}, {Schellart}, {Shaw}, {Silva}, {Shiokawa},
  {Smith}, {Snow}, {Souccar}, {Sousa}, {Sridharan}, {Srinivasan}, {Stahm},
  {Stark}, {Story}, {Timmer}, {Vertatschitsch}, {Walther}, {Wei}, {Whitehorn},
  {Whitney}, {Woody}, {Wouterloot}, {Wright}, {Yamaguchi}, {Yu}, {Zeballos},
  {Zhang}, \& {Ziurys}}]{EHT2019a}
{Event Horizon Telescope Collaboration}, {Akiyama}, K., {Alberdi}, A., {et~al.}
  2019{\natexlab{a}}, \apjl, 875, L1, \dodoi{10.3847/2041-8213/ab0ec7}

\bibitem[{{Event Horizon Telescope Collaboration}
  {et~al.}(2019{\natexlab{b}}){Event Horizon Telescope Collaboration},
  {Akiyama}, {Alberdi}, {Alef}, {Asada}, {Azulay}, {Baczko}, {Ball},
  {Balokovi{\'c}}, {Barrett}, {Bintley}, {Blackburn}, {Boland}, {Bouman},
  {Bower}, {Bremer}, {Brinkerink}, {Brissenden}, {Britzen}, {Broderick},
  {Broguiere}, {Bronzwaer}, {Byun}, {Carlstrom}, {Chael}, {Chan}, {Chatterjee},
  {Chatterjee}, {Chen}, {Chen}, {Cho}, {Christian}, {Conway}, {Cordes}, {Crew},
  {Cui}, {Davelaar}, {De Laurentis}, {Deane}, {Dempsey}, {Desvignes}, {Dexter},
  {Doeleman}, {Eatough}, {Falcke}, {Fish}, {Fomalont}, {Fraga-Encinas},
  {Friberg}, {Fromm}, {G{\'o}mez}, {Galison}, {Gammie}, {Garc{\'\i}a},
  {Gentaz}, {Georgiev}, {Goddi}, {Gold}, {Gu}, {Gurwell}, {Hada}, {Hecht},
  {Hesper}, {Ho}, {Ho}, {Honma}, {Huang}, {Huang}, {Hughes}, {Ikeda}, {Inoue},
  {Issaoun}, {James}, {Jannuzi}, {Janssen}, {Jeter}, {Jiang}, {Johnson},
  {Jorstad}, {Jung}, {Karami}, {Karuppusamy}, {Kawashima}, {Keating},
  {Kettenis}, {Kim}, {Kim}, {Kim}, {Kino}, {Koay}, {Koch}, {Koyama}, {Kramer},
  {Kramer}, {Krichbaum}, {Kuo}, {Lauer}, {Lee}, {Li}, {Li}, {Lindqvist}, {Liu},
  {Liuzzo}, {Lo}, {Lobanov}, {Loinard}, {Lonsdale}, {Lu}, {MacDonald}, {Mao},
  {Markoff}, {Marrone}, {Marscher}, {Mart{\'\i}-Vidal}, {Matsushita},
  {Matthews}, {Medeiros}, {Menten}, {Mizuno}, {Mizuno}, {Moran}, {Moriyama},
  {Moscibrodzka}, {M{\"u}ller}, {Nagai}, {Nagar}, {Nakamura}, {Narayan},
  {Narayanan}, {Natarajan}, {Neri}, {Ni}, {Noutsos}, {Okino}, {Olivares},
  {Ortiz-Le{\'o}n}, {Oyama}, {{\"O}zel}, {Palumbo}, {Patel}, {Pen}, {Pesce},
  {Pi{\'e}tu}, {Plambeck}, {PopStefanija}, {Porth}, {Prather},
  {Preciado-L{\'o}pez}, {Psaltis}, {Pu}, {Ramakrishnan}, {Rao}, {Rawlings},
  {Raymond}, {Rezzolla}, {Ripperda}, {Roelofs}, {Rogers}, {Ros}, {Rose},
  {Roshanineshat}, {Rottmann}, {Roy}, {Ruszczyk}, {Ryan}, {Rygl},
  {S{\'a}nchez}, {S{\'a}nchez-Arguelles}, {Sasada}, {Savolainen}, {Schloerb},
  {Schuster}, {Shao}, {Shen}, {Small}, {Sohn}, {SooHoo}, {Tazaki}, {Tiede},
  {Tilanus}, {Titus}, {Toma}, {Torne}, {Trent}, {Trippe}, {Tsuda}, {van
  Bemmel}, {van Langevelde}, {van Rossum}, {Wagner}, {Wardle}, {Weintroub},
  {Wex}, {Wharton}, {Wielgus}, {Wong}, {Wu}, {Young}, {Young}, {Younsi},
  {Yuan}, {Yuan}, {Zensus}, {Zhao}, {Zhao}, {Zhu}, {Algaba}, {Allardi},
  {Amestica}, {Bach}, {Beaudoin}, {Benson}, {Berthold}, {Blanchard},
  {Blundell}, {Bustamente}, {Cappallo}, {Castillo-Dom{\'\i}nguez}, {Chang},
  {Chang}, {Chang}, {Chen}, {Chilson}, {Chuter}, {C{\'o}rdova Rosado},
  {Coulson}, {Crawford}, {Crowley}, {David}, {Derome}, {Dexter}, {Dornbusch},
  {Dudevoir}, {Dzib}, {Eckert}, {Erickson}, {Everett}, {Faber}, {Farah},
  {Fath}, {Folkers}, {Forbes}, {Freund}, {G{\'o}mez-Ruiz}, {Gale}, {Gao},
  {Geertsema}, {Graham}, {Greer}, {Grosslein}, {Gueth}, {Halverson}, {Han},
  {Han}, {Hao}, {Hasegawa}, {Henning}, {Hern{\'a}ndez-G{\'o}mez},
  {Herrero-Illana}, {Heyminck}, {Hirota}, {Hoge}, {Huang}, {Impellizzeri},
  {Jiang}, {Kamble}, {Keisler}, {Kimura}, {Kono}, {Kubo}, {Kuroda}, {Lacasse},
  {Laing}, {Leitch}, {Li}, {Lin}, {Liu}, {Liu}, {Lu}, {Marson},
  {Martin-Cocher}, {Massingill}, {Matulonis}, {McColl}, {McWhirter}, {Messias},
  {Meyer-Zhao}, {Michalik}, {Monta{\~n}a}, {Montgomerie}, {Mora-Klein},
  {Muders}, {Nadolski}, {Navarro}, {Nguyen}, {Nishioka}, {Norton}, {Nystrom},
  {Ogawa}, {Oshiro}, {Oyama}, {Padin}, {Parsons}, {Paine}, {Pe{\~n}alver},
  {Phillips}, {Poirier}, {Pradel}, {Primiani}, {Raffin}, {Rahlin}, {Reiland},
  {Risacher}, {Ruiz}, {S{\'a}ez-Mada{\'\i}n}, {Sassella}, {Schellart}, {Shaw},
  {Silva}, {Shiokawa}, {Smith}, {Snow}, {Souccar}, {Sousa}, {Sridharan},
  {Srinivasan}, {Stahm}, {Stark}, {Story}, {Timmer}, {Vertatschitsch},
  {Walther}, {Wei}, {Whitehorn}, {Whitney}, {Woody}, {Wouterloot}, {Wright},
  {Yamaguchi}, {Yu}, {Zeballos}, \& {Ziurys}}]{EHT2019b}
---. 2019{\natexlab{b}}, \apjl, 875, L2, \dodoi{10.3847/2041-8213/ab0c96}

\bibitem[{{Event Horizon Telescope Collaboration}
  {et~al.}(2019{\natexlab{c}}){Event Horizon Telescope Collaboration},
  {Akiyama}, {Alberdi}, {Alef}, {Asada}, {Azulay}, {Baczko}, {Ball},
  {Balokovi{\'c}}, {Barrett}, {Bintley}, {Blackburn}, {Boland}, {Bouman},
  {Bower}, {Bremer}, {Brinkerink}, {Brissenden}, {Britzen}, {Broderick},
  {Broguiere}, {Bronzwaer}, {Byun}, {Carlstrom}, {Chael}, {Chan}, {Chatterjee},
  {Chatterjee}, {Chen}, {Chen}, {Cho}, {Christian}, {Conway}, {Cordes}, {Crew},
  {Cui}, {Davelaar}, {De Laurentis}, {Deane}, {Dempsey}, {Desvignes}, {Dexter},
  {Doeleman}, {Eatough}, {Falcke}, {Fish}, {Fomalont}, {Fraga-Encinas},
  {Friberg}, {Fromm}, {G{\'o}mez}, {Galison}, {Gammie}, {Garc{\'\i}a},
  {Gentaz}, {Georgiev}, {Goddi}, {Gold}, {Gu}, {Gurwell}, {Hada}, {Hecht},
  {Hesper}, {Ho}, {Ho}, {Honma}, {Huang}, {Huang}, {Hughes}, {Ikeda}, {Inoue},
  {Issaoun}, {James}, {Jannuzi}, {Janssen}, {Jeter}, {Jiang}, {Johnson},
  {Jorstad}, {Jung}, {Karami}, {Karuppusamy}, {Kawashima}, {Keating},
  {Kettenis}, {Kim}, {Kim}, {Kim}, {Kino}, {Koay}, {Koch}, {Koyama}, {Kramer},
  {Kramer}, {Krichbaum}, {Kuo}, {Lauer}, {Lee}, {Li}, {Li}, {Lindqvist}, {Liu},
  {Liuzzo}, {Lo}, {Lobanov}, {Loinard}, {Lonsdale}, {Lu}, {MacDonald}, {Mao},
  {Markoff}, {Marrone}, {Marscher}, {Mart{\'\i}-Vidal}, {Matsushita},
  {Matthews}, {Medeiros}, {Menten}, {Mizuno}, {Mizuno}, {Moran}, {Moriyama},
  {Moscibrodzka}, {M{\"u}ller}, {Nagai}, {Nagar}, {Nakamura}, {Narayan},
  {Narayanan}, {Natarajan}, {Neri}, {Ni}, {Noutsos}, {Okino}, {Olivares},
  {Ortiz-Le{\'o}n}, {Oyama}, {{\"O}zel}, {Palumbo}, {Patel}, {Pen}, {Pesce},
  {Pi{\'e}tu}, {Plambeck}, {PopStefanija}, {Porth}, {Prather},
  {Preciado-L{\'o}pez}, {Psaltis}, {Pu}, {Ramakrishnan}, {Rao}, {Rawlings},
  {Raymond}, {Rezzolla}, {Ripperda}, {Roelofs}, {Rogers}, {Ros}, {Rose},
  {Roshanineshat}, {Rottmann}, {Roy}, {Ruszczyk}, {Ryan}, {Rygl},
  {S{\'a}nchez}, {S{\'a}nchez-Arguelles}, {Sasada}, {Savolainen}, {Schloerb},
  {Schuster}, {Shao}, {Shen}, {Small}, {Sohn}, {SooHoo}, {Tazaki}, {Tiede},
  {Tilanus}, {Titus}, {Toma}, {Torne}, {Trent}, {Trippe}, {Tsuda}, {van
  Bemmel}, {van Langevelde}, {van Rossum}, {Wagner}, {Wardle}, {Weintroub},
  {Wex}, {Wharton}, {Wielgus}, {Wong}, {Wu}, {Young}, {Young}, {Younsi},
  {Yuan}, {Yuan}, {Zensus}, {Zhao}, {Zhao}, {Zhu}, {Cappallo}, {Farah},
  {Folkers}, {Meyer-Zhao}, {Michalik}, {Nadolski}, {Nishioka}, {Pradel},
  {Primiani}, {Souccar}, {Vertatschitsch}, \& {Yamaguchi}}]{EHT2019c}
---. 2019{\natexlab{c}}, \apjl, 875, L3, \dodoi{10.3847/2041-8213/ab0c57}

\bibitem[{{Event Horizon Telescope Collaboration}
  {et~al.}(2019{\natexlab{d}}){Event Horizon Telescope Collaboration},
  {Akiyama}, {Alberdi}, {Alef}, {Asada}, {Azulay}, {Baczko}, {Ball},
  {Balokovi{\'c}}, {Barrett}, {Bintley}, {Blackburn}, {Boland}, {Bouman},
  {Bower}, {Bremer}, {Brinkerink}, {Brissenden}, {Britzen}, {Broderick},
  {Broguiere}, {Bronzwaer}, {Byun}, {Carlstrom}, {Chael}, {Chan}, {Chatterjee},
  {Chatterjee}, {Chen}, {Chen}, {Cho}, {Christian}, {Conway}, {Cordes}, {Crew},
  {Cui}, {Davelaar}, {De Laurentis}, {Deane}, {Dempsey}, {Desvignes}, {Dexter},
  {Doeleman}, {Eatough}, {Falcke}, {Fish}, {Fomalont}, {Fraga-Encinas},
  {Freeman}, {Friberg}, {Fromm}, {G{\'o}mez}, {Galison}, {Gammie},
  {Garc{\'\i}a}, {Gentaz}, {Georgiev}, {Goddi}, {Gold}, {Gu}, {Gurwell},
  {Hada}, {Hecht}, {Hesper}, {Ho}, {Ho}, {Honma}, {Huang}, {Huang}, {Hughes},
  {Ikeda}, {Inoue}, {Issaoun}, {James}, {Jannuzi}, {Janssen}, {Jeter}, {Jiang},
  {Johnson}, {Jorstad}, {Jung}, {Karami}, {Karuppusamy}, {Kawashima},
  {Keating}, {Kettenis}, {Kim}, {Kim}, {Kim}, {Kino}, {Koay}, {Koch}, {Koyama},
  {Kramer}, {Kramer}, {Krichbaum}, {Kuo}, {Lauer}, {Lee}, {Li}, {Li},
  {Lindqvist}, {Liu}, {Liuzzo}, {Lo}, {Lobanov}, {Loinard}, {Lonsdale}, {Lu},
  {MacDonald}, {Mao}, {Markoff}, {Marrone}, {Marscher}, {Mart{\'\i}-Vidal},
  {Matsushita}, {Matthews}, {Medeiros}, {Menten}, {Mizuno}, {Mizuno}, {Moran},
  {Moriyama}, {Moscibrodzka}, {M{\"u}ller}, {Nagai}, {Nagar}, {Nakamura},
  {Narayan}, {Narayanan}, {Natarajan}, {Neri}, {Ni}, {Noutsos}, {Okino},
  {Olivares}, {Oyama}, {{\"O}zel}, {Palumbo}, {Patel}, {Pen}, {Pesce},
  {Pi{\'e}tu}, {Plambeck}, {PopStefanija}, {Porth}, {Prather},
  {Preciado-L{\'o}pez}, {Psaltis}, {Pu}, {Ramakrishnan}, {Rao}, {Rawlings},
  {Raymond}, {Rezzolla}, {Ripperda}, {Roelofs}, {Rogers}, {Ros}, {Rose},
  {Roshanineshat}, {Rottmann}, {Roy}, {Ruszczyk}, {Ryan}, {Rygl},
  {S{\'a}nchez}, {S{\'a}nchez-Arguelles}, {Sasada}, {Savolainen}, {Schloerb},
  {Schuster}, {Shao}, {Shen}, {Small}, {Sohn}, {SooHoo}, {Tazaki}, {Tiede},
  {Tilanus}, {Titus}, {Toma}, {Torne}, {Trent}, {Trippe}, {Tsuda}, {van
  Bemmel}, {van Langevelde}, {van Rossum}, {Wagner}, {Wardle}, {Weintroub},
  {Wex}, {Wharton}, {Wielgus}, {Wong}, {Wu}, {Young}, {Young}, {Younsi},
  {Yuan}, {Yuan}, {Zensus}, {Zhao}, {Zhao}, {Zhu}, {Farah}, {Meyer-Zhao},
  {Michalik}, {Nadolski}, {Nishioka}, {Pradel}, {Primiani}, {Souccar},
  {Vertatschitsch}, \& {Yamaguchi}}]{EHT2019d}
---. 2019{\natexlab{d}}, \apjl, 875, L4, \dodoi{10.3847/2041-8213/ab0e85}

\bibitem[{{Event Horizon Telescope Collaboration}
  {et~al.}(2019{\natexlab{e}}){Event Horizon Telescope Collaboration},
  {Akiyama}, {Alberdi}, {Alef}, {Asada}, {Azulay}, {Baczko}, {Ball},
  {Balokovi{\'c}}, {Barrett}, {Bintley}, {Blackburn}, {Boland}, {Bouman},
  {Bower}, {Bremer}, {Brinkerink}, {Brissenden}, {Britzen}, {Broderick},
  {Broguiere}, {Bronzwaer}, {Byun}, {Carlstrom}, {Chael}, {Chan}, {Chatterjee},
  {Chatterjee}, {Chen}, {Chen}, {Cho}, {Christian}, {Conway}, {Cordes}, {Crew},
  {Cui}, {Davelaar}, {De Laurentis}, {Deane}, {Dempsey}, {Desvignes}, {Dexter},
  {Doeleman}, {Eatough}, {Falcke}, {Fish}, {Fomalont}, {Fraga-Encinas},
  {Friberg}, {Fromm}, {G{\'o}mez}, {Galison}, {Gammie}, {Garc{\'\i}a},
  {Gentaz}, {Georgiev}, {Goddi}, {Gold}, {Gu}, {Gurwell}, {Hada}, {Hecht},
  {Hesper}, {Ho}, {Ho}, {Honma}, {Huang}, {Huang}, {Hughes}, {Ikeda}, {Inoue},
  {Issaoun}, {James}, {Jannuzi}, {Janssen}, {Jeter}, {Jiang}, {Johnson},
  {Jorstad}, {Jung}, {Karami}, {Karuppusamy}, {Kawashima}, {Keating},
  {Kettenis}, {Kim}, {Kim}, {Kim}, {Kino}, {Koay}, {Koch}, {Koyama}, {Kramer},
  {Kramer}, {Krichbaum}, {Kuo}, {Lauer}, {Lee}, {Li}, {Li}, {Lindqvist}, {Liu},
  {Liuzzo}, {Lo}, {Lobanov}, {Loinard}, {Lonsdale}, {Lu}, {MacDonald}, {Mao},
  {Markoff}, {Marrone}, {Marscher}, {Mart{\'\i}-Vidal}, {Matsushita},
  {Matthews}, {Medeiros}, {Menten}, {Mizuno}, {Mizuno}, {Moran}, {Moriyama},
  {Moscibrodzka}, {Mul{\ensuremath{\ddot{}}}ler}, {Nagai}, {Nagar}, {Nakamura},
  {Narayan}, {Narayanan}, {Natarajan}, {Neri}, {Ni}, {Noutsos}, {Okino},
  {Olivares}, {Oyama}, {{\"O}zel}, {Palumbo}, {Patel}, {Pen}, {Pesce},
  {Pi{\'e}tu}, {Plambeck}, {PopStefanija}, {Porth}, {Prather},
  {Preciado-L{\'o}pez}, {Psaltis}, {Pu}, {Ramakrishnan}, {Rao}, {Rawlings},
  {Raymond}, {Rezzolla}, {Ripperda}, {Roelofs}, {Rogers}, {Ros}, {Rose},
  {Roshanineshat}, {Rottmann}, {Roy}, {Ruszczyk}, {Ryan}, {Rygl},
  {S{\'a}nchez}, {S{\'a}nchez-Arguelles}, {Sasada}, {Savolainen}, {Schloerb},
  {Schuster}, {Shao}, {Shen}, {Small}, {Sohn}, {SooHoo}, {Tazaki}, {Tiede},
  {Tilanus}, {Titus}, {Toma}, {Torne}, {Trent}, {Trippe}, {Tsuda}, {van
  Bemmel}, {van Langevelde}, {van Rossum}, {Wagner}, {Wardle}, {Weintroub},
  {Wex}, {Wharton}, {Wielgus}, {Wong}, {Wu}, {Young}, {Young}, {Younsi},
  {Yuan}, {Yuan}, {Zensus}, {Zhao}, {Zhao}, {Zhu}, {Anczarski}, {Baganoff},
  {Eckart}, {Farah}, {Haggard}, {Meyer-Zhao}, {Michalik}, {Nadolski},
  {Neilsen}, {Nishioka}, {Nowak}, {Pradel}, {Primiani}, {Souccar},
  {Vertatschitsch}, {Yamaguchi}, \& {Zhang}}]{EHT2019e}
---. 2019{\natexlab{e}}, \apjl, 875, L5, \dodoi{10.3847/2041-8213/ab0f43}

\bibitem[{{Event Horizon Telescope Collaboration}
  {et~al.}(2019{\natexlab{f}}){Event Horizon Telescope Collaboration},
  {Akiyama}, {Alberdi}, {Alef}, {Asada}, {Azulay}, {Baczko}, {Ball},
  {Balokovi{\'c}}, {Barrett}, {Bintley}, {Blackburn}, {Boland}, {Bouman},
  {Bower}, {Bremer}, {Brinkerink}, {Brissenden}, {Britzen}, {Broderick},
  {Broguiere}, {Bronzwaer}, {Byun}, {Carlstrom}, {Chael}, {Chan}, {Chatterjee},
  {Chatterjee}, {Chen}, {Chen}, {Cho}, {Christian}, {Conway}, {Cordes}, {Crew},
  {Cui}, {Davelaar}, {De Laurentis}, {Deane}, {Dempsey}, {Desvignes}, {Dexter},
  {Doeleman}, {Eatough}, {Falcke}, {Fish}, {Fomalont}, {Fraga-Encinas},
  {Friberg}, {Fromm}, {G{\'o}mez}, {Galison}, {Gammie}, {Garc{\'\i}a},
  {Gentaz}, {Georgiev}, {Goddi}, {Gold}, {Gu}, {Gurwell}, {Hada}, {Hecht},
  {Hesper}, {Ho}, {Ho}, {Honma}, {Huang}, {Huang}, {Hughes}, {Ikeda}, {Inoue},
  {Issaoun}, {James}, {Jannuzi}, {Janssen}, {Jeter}, {Jiang}, {Johnson},
  {Jorstad}, {Jung}, {Karami}, {Karuppusamy}, {Kawashima}, {Keating},
  {Kettenis}, {Kim}, {Kim}, {Kim}, {Kino}, {Koay}, {Koch}, {Koyama}, {Kramer},
  {Kramer}, {Krichbaum}, {Kuo}, {Lauer}, {Lee}, {Li}, {Li}, {Lindqvist}, {Liu},
  {Liuzzo}, {Lo}, {Lobanov}, {Loinard}, {Lonsdale}, {Lu}, {MacDonald}, {Mao},
  {Markoff}, {Marrone}, {Marscher}, {Mart{\'\i}-Vidal}, {Matsushita},
  {Matthews}, {Medeiros}, {Menten}, {Mizuno}, {Mizuno}, {Moran}, {Moriyama},
  {Moscibrodzka}, {M{\"u}ller}, {Nagai}, {Nagar}, {Nakamura}, {Narayan},
  {Narayanan}, {Natarajan}, {Neri}, {Ni}, {Noutsos}, {Okino}, {Olivares},
  {Oyama}, {{\"O}zel}, {Palumbo}, {Patel}, {Pen}, {Pesce}, {Pi{\'e}tu},
  {Plambeck}, {PopStefanija}, {Porth}, {Prather}, {Preciado-L{\'o}pez},
  {Psaltis}, {Pu}, {Ramakrishnan}, {Rao}, {Rawlings}, {Raymond}, {Rezzolla},
  {Ripperda}, {Roelofs}, {Rogers}, {Ros}, {Rose}, {Roshanineshat}, {Rottmann},
  {Roy}, {Ruszczyk}, {Ryan}, {Rygl}, {S{\'a}nchez}, {S{\'a}nchez-Arguelles},
  {Sasada}, {Savolainen}, {Schloerb}, {Schuster}, {Shao}, {Shen}, {Small},
  {Sohn}, {SooHoo}, {Tazaki}, {Tiede}, {Tilanus}, {Titus}, {Toma}, {Torne},
  {Trent}, {Trippe}, {Tsuda}, {van Bemmel}, {van Langevelde}, {van Rossum},
  {Wagner}, {Wardle}, {Weintroub}, {Wex}, {Wharton}, {Wielgus}, {Wong}, {Wu},
  {Young}, {Young}, {Younsi}, {Yuan}, {Yuan}, {Zensus}, {Zhao}, {Zhao}, {Zhu},
  {Farah}, {Meyer-Zhao}, {Michalik}, {Nadolski}, {Nishioka}, {Pradel},
  {Primiani}, {Souccar}, {Vertatschitsch}, \& {Yamaguchi}}]{EHT2019f}
---. 2019{\natexlab{f}}, \apjl, 875, L6, \dodoi{10.3847/2041-8213/ab1141}

\bibitem[{{Event Horizon Telescope Collaboration}
  {et~al.}(2021{\natexlab{a}}){Event Horizon Telescope Collaboration},
  {Akiyama}, {Algaba}, {Alberdi}, {Alef}, {Anantua}, {Asada}, {Azulay},
  {Baczko}, {Ball}, {Balokovi{\'c}}, {Barrett}, {Benson}, {Bintley},
  {Blackburn}, {Blundell}, {Boland}, {Bouman}, {Bower}, {Boyce}, {Bremer},
  {Brinkerink}, {Brissenden}, {Britzen}, {Broderick}, {Broguiere}, {Bronzwaer},
  {Byun}, {Carlstrom}, {Chael}, {Chan}, {Chatterjee}, {Chatterjee}, {Chen},
  {Chen}, {Chesler}, {Cho}, {Christian}, {Conway}, {Cordes}, {Crawford},
  {Crew}, {Cruz-Osorio}, {Cui}, {Davelaar}, {De Laurentis}, {Deane}, {Dempsey},
  {Desvignes}, {Dexter}, {Doeleman}, {Eatough}, {Falcke}, {Farah}, {Fish},
  {Fomalont}, {Ford}, {Fraga-Encinas}, {Freeman}, {Friberg}, {Fromm},
  {Fuentes}, {Galison}, {Gammie}, {Garc{\'\i}a}, {Gentaz}, {Georgiev}, {Goddi},
  {Gold}, {G{\'o}mez}, {G{\'o}mez-Ruiz}, {Gu}, {Gurwell}, {Hada}, {Haggard},
  {Hecht}, {Hesper}, {Ho}, {Ho}, {Honma}, {Huang}, {Huang}, {Hughes}, {Ikeda},
  {Inoue}, {Issaoun}, {James}, {Jannuzi}, {Janssen}, {Jeter}, {Jiang},
  {Jimenez-Rosales}, {Johnson}, {Jorstad}, {Jung}, {Karami}, {Karuppusamy},
  {Kawashima}, {Keating}, {Kettenis}, {Kim}, {Kim}, {Kim}, {Kim}, {Kino},
  {Koay}, {Kofuji}, {Koch}, {Koyama}, {Kramer}, {Kramer}, {Krichbaum}, {Kuo},
  {Lauer}, {Lee}, {Levis}, {Li}, {Li}, {Lindqvist}, {Lico}, {Lindahl}, {Liu},
  {Liu}, {Liuzzo}, {Lo}, {Lobanov}, {Loinard}, {Lonsdale}, {Lu}, {MacDonald},
  {Mao}, {Marchili}, {Markoff}, {Marrone}, {Marscher}, {Mart{\'\i}-Vidal},
  {Matsushita}, {Matthews}, {Medeiros}, {Menten}, {Mizuno}, {Mizuno}, {Moran},
  {Moriyama}, {Moscibrodzka}, {M{\"u}ller}, {Musoke}, {Mej{\'\i}as},
  {Michalik}, {Nadolski}, {Nagai}, {Nagar}, {Nakamura}, {Narayan}, {Narayanan},
  {Natarajan}, {Nathanail}, {Neilsen}, {Neri}, {Ni}, {Noutsos}, {Nowak},
  {Okino}, {Olivares}, {Ortiz-Le{\'o}n}, {Oyama}, {{\"O}zel}, {Palumbo},
  {Park}, {Patel}, {Pen}, {Pesce}, {Pi{\'e}tu}, {Plambeck}, {PopStefanija},
  {Porth}, {P{\"o}tzl}, {Prather}, {Preciado-L{\'o}pez}, {Psaltis}, {Pu},
  {Ramakrishnan}, {Rao}, {Rawlings}, {Raymond}, {Rezzolla}, {Ricarte},
  {Ripperda}, {Roelofs}, {Rogers}, {Ros}, {Rose}, {Roshanineshat}, {Rottmann},
  {Roy}, {Ruszczyk}, {Rygl}, {S{\'a}nchez}, {S{\'a}nchez-Arguelles}, {Sasada},
  {Savolainen}, {Schloerb}, {Schuster}, {Shao}, {Shen}, {Small}, {Sohn},
  {SooHoo}, {Sun}, {Tazaki}, {Tetarenko}, {Tiede}, {Tilanus}, {Titus}, {Toma},
  {Torne}, {Trent}, {Traianou}, {Trippe}, {van Bemmel}, {van Langevelde}, {van
  Rossum}, {Wagner}, {Ward-Thompson}, {Wardle}, {Weintroub}, {Wex}, {Wharton},
  {Wielgus}, {Wong}, {Wu}, {Yoon}, {Young}, {Young}, {Younsi}, {Yuan}, {Yuan},
  {Zensus}, {Zhao}, \& {Zhao}}]{EHT2021a}
{Event Horizon Telescope Collaboration}, {Akiyama}, K., {Algaba}, J.~C.,
  {et~al.} 2021{\natexlab{a}}, \apjl, 910, L12,
  \dodoi{10.3847/2041-8213/abe71d}

\bibitem[{{Event Horizon Telescope Collaboration}
  {et~al.}(2021{\natexlab{b}}){Event Horizon Telescope Collaboration},
  {Akiyama}, {Algaba}, {Alberdi}, {Alef}, {Anantua}, {Asada}, {Azulay},
  {Baczko}, {Ball}, {Balokovi{\'c}}, {Barrett}, {Benson}, {Bintley},
  {Blackburn}, {Blundell}, {Boland}, {Bouman}, {Bower}, {Boyce}, {Bremer},
  {Brinkerink}, {Brissenden}, {Britzen}, {Broderick}, {Broguiere}, {Bronzwaer},
  {Byun}, {Carlstrom}, {Chael}, {Chan}, {Chatterjee}, {Chatterjee}, {Chen},
  {Chen}, {Chesler}, {Cho}, {Christian}, {Conway}, {Cordes}, {Crawford},
  {Crew}, {Cruz-Osorio}, {Cui}, {Davelaar}, {De Laurentis}, {Deane}, {Dempsey},
  {Desvignes}, {Dexter}, {Doeleman}, {Eatough}, {Falcke}, {Farah}, {Fish},
  {Fomalont}, {Ford}, {Fraga-Encinas}, {Friberg}, {Fromm}, {Fuentes},
  {Galison}, {Gammie}, {Garc{\'\i}a}, {Gelles}, {Gentaz}, {Georgiev}, {Goddi},
  {Gold}, {G{\'o}mez}, {G{\'o}mez-Ruiz}, {Gu}, {Gurwell}, {Hada}, {Haggard},
  {Hecht}, {Hesper}, {Himwich}, {Ho}, {Ho}, {Honma}, {Huang}, {Huang},
  {Hughes}, {Ikeda}, {Inoue}, {Issaoun}, {James}, {Jannuzi}, {Janssen},
  {Jeter}, {Jiang}, {Jimenez-Rosales}, {Johnson}, {Jorstad}, {Jung}, {Karami},
  {Karuppusamy}, {Kawashima}, {Keating}, {Kettenis}, {Kim}, {Kim}, {Kim},
  {Kim}, {Kino}, {Koay}, {Kofuji}, {Koch}, {Koyama}, {Kramer}, {Kramer},
  {Krichbaum}, {Kuo}, {Lauer}, {Lee}, {Levis}, {Li}, {Li}, {Lindqvist}, {Lico},
  {Lindahl}, {Liu}, {Liu}, {Liuzzo}, {Lo}, {Lobanov}, {Loinard}, {Lonsdale},
  {Lu}, {MacDonald}, {Mao}, {Marchili}, {Markoff}, {Marrone}, {Marscher},
  {Mart{\'\i}-Vidal}, {Matsushita}, {Matthews}, {Medeiros}, {Menten}, {Mizuno},
  {Mizuno}, {Moran}, {Moriyama}, {Moscibrodzka}, {M{\"u}ller}, {Musoke}, {Mus
  Mej{\'\i}as}, {Michalik}, {Nadolski}, {Nagai}, {Nagar}, {Nakamura},
  {Narayan}, {Narayanan}, {Natarajan}, {Nathanail}, {Neilsen}, {Neri}, {Ni},
  {Noutsos}, {Nowak}, {Okino}, {Olivares}, {Ortiz-Le{\'o}n}, {Oyama},
  {{\"O}zel}, {Palumbo}, {Park}, {Patel}, {Pen}, {Pesce}, {Pi{\'e}tu},
  {Plambeck}, {PopStefanija}, {Porth}, {P{\"o}tzl}, {Prather},
  {Preciado-L{\'o}pez}, {Psaltis}, {Pu}, {Ramakrishnan}, {Rao}, {Rawlings},
  {Raymond}, {Rezzolla}, {Ricarte}, {Ripperda}, {Roelofs}, {Rogers}, {Ros},
  {Rose}, {Roshanineshat}, {Rottmann}, {Roy}, {Ruszczyk}, {Rygl},
  {S{\'a}nchez}, {S{\'a}nchez-Arguelles}, {Sasada}, {Savolainen}, {Schloerb},
  {Schuster}, {Shao}, {Shen}, {Small}, {Sohn}, {SooHoo}, {Sun}, {Tazaki},
  {Tetarenko}, {Tiede}, {Tilanus}, {Titus}, {Toma}, {Torne}, {Trent},
  {Traianou}, {Trippe}, {van Bemmel}, {van Langevelde}, {van Rossum}, {Wagner},
  {Ward-Thompson}, {Wardle}, {Weintroub}, {Wex}, {Wharton}, {Wielgus}, {Wong},
  {Wu}, {Yoon}, {Young}, {Young}, {Younsi}, {Yuan}, {Yuan}, {Zensus}, {Zhao},
  \& {Zhao}}]{EHT2021b}
---. 2021{\natexlab{b}}, \apjl, 910, L13, \dodoi{10.3847/2041-8213/abe4de}

\bibitem[{{Gabuzda}(2018)}]{Gabuzda2018}
{Gabuzda}, D. 2018, Galaxies, 7, 5, \dodoi{10.3390/galaxies7010005}

\bibitem[{{G{\'o}mez} {et~al.}(2022){G{\'o}mez}, {Traianou}, {Krichbaum},
  {Lobanov}, {Fuentes}, {Lico}, {Zhao}, {Bruni}, {Kovalev},
  {L{\"a}hteenm{\"a}ki}, {Voitsik}, {Lisakov}, {Angelakis}, {Bach}, {Casadio},
  {Cho}, {Dey}, {Gopakumar}, {Gurvits}, {Jorstad}, {Kovalev}, {Lister},
  {Marscher}, {Myserlis}, {Pushkarev}, {Ros}, {Savolainen}, {Tornikoski},
  {Valtonen}, \& {Zensus}}]{Gomez2022}
{G{\'o}mez}, J.~L., {Traianou}, E., {Krichbaum}, T.~P., {et~al.} 2022, \apj,
  924, 122, \dodoi{10.3847/1538-4357/ac3bcc}

\bibitem[{{Greisen}(2003)}]{Greisen2003}
{Greisen}, E.~W. 2003, Astrophysics and Space Science Library, Vol. 285, {AIPS,
  the VLA, and the VLBA}, ed. A.~{Heck}, 109, \dodoi{10.1007/0-306-48080-8_7}

\bibitem[{{Hada} {et~al.}(2016){Hada}, {Kino}, {Doi}, {Nagai}, {Honma},
  {Akiyama}, {Tazaki}, {Lico}, {Giroletti}, {Giovannini}, {Orienti}, \&
  {Hagiwara}}]{Hada2016}
{Hada}, K., {Kino}, M., {Doi}, A., {et~al.} 2016, \apj, 817, 131,
  \dodoi{10.3847/0004-637X/817/2/131}

\bibitem[{{Hamaker}(2000)}]{Hamaker2000}
{Hamaker}, J.~P. 2000, \aaps, 143, 515, \dodoi{10.1051/aas:2000337}

\bibitem[{{Hamaker} \& {Bregman}(1996)}]{HB1996}
{Hamaker}, J.~P., \& {Bregman}, J.~D. 1996, \aaps, 117, 161

\bibitem[{{Hamaker} {et~al.}(1996){Hamaker}, {Bregman}, \&
  {Sault}}]{Hamaker1996}
{Hamaker}, J.~P., {Bregman}, J.~D., \& {Sault}, R.~J. 1996, \aaps, 117, 137

\bibitem[{{Hovatta} {et~al.}(2012){Hovatta}, {Lister}, {Aller}, {Aller},
  {Homan}, {Kovalev}, {Pushkarev}, \& {Savolainen}}]{Hovatta2012}
{Hovatta}, T., {Lister}, M.~L., {Aller}, M.~F., {et~al.} 2012, \aj, 144, 105,
  \dodoi{10.1088/0004-6256/144/4/105}

\bibitem[{{Jones}(1941)}]{Jones1941}
{Jones}, R.~C. 1941, Journal of the Optical Society of America (1917-1983), 31,
  488

\bibitem[{{Jorstad} {et~al.}(2017){Jorstad}, {Marscher}, {Morozova},
  {Troitsky}, {Agudo}, {Casadio}, {Foord}, {G{\'o}mez}, {MacDonald}, {Molina},
  {L{\"a}hteenm{\"a}ki}, {Tammi}, \& {Tornikoski}}]{Jorstad2017}
{Jorstad}, S.~G., {Marscher}, A.~P., {Morozova}, D.~A., {et~al.} 2017, \apj,
  846, 98, \dodoi{10.3847/1538-4357/aa8407}

\bibitem[{{Kettenis} {et~al.}(2006){Kettenis}, {van Langevelde}, {Reynolds}, \&
  {Cotton}}]{Kettenis2006}
{Kettenis}, M., {van Langevelde}, H.~J., {Reynolds}, C., \& {Cotton}, B. 2006,
  in Astronomical Society of the Pacific Conference Series, Vol. 351,
  Astronomical Data Analysis Software and Systems XV, ed. C.~{Gabriel},
  C.~{Arviset}, D.~{Ponz}, \& S.~{Enrique}, 497

\bibitem[{{Kim} {et~al.}(2019){Kim}, {Krichbaum}, {Marscher}, {Jorstad},
  {Agudo}, {Thum}, {Hodgson}, {MacDonald}, {Ros}, {Lu}, {Bremer}, {de Vicente},
  {Lindqvist}, {Trippe}, \& {Zensus}}]{Kim2019}
{Kim}, J.~Y., {Krichbaum}, T.~P., {Marscher}, A.~P., {et~al.} 2019, \aap, 622,
  A196, \dodoi{10.1051/0004-6361/201832920}

\bibitem[{{Leppanen} {et~al.}(1995){Leppanen}, {Zensus}, \&
  {Diamond}}]{Leppanen1995}
{Leppanen}, K.~J., {Zensus}, J.~A., \& {Diamond}, P.~J. 1995, \aj, 110, 2479,
  \dodoi{10.1086/117706}

\bibitem[{{Lisakov} {et~al.}(2021){Lisakov}, {Kravchenko}, {Pushkarev},
  {Kovalev}, {Savolainen}, \& {Lister}}]{Lisakov2021}
{Lisakov}, M.~M., {Kravchenko}, E.~V., {Pushkarev}, A.~B., {et~al.} 2021, \apj,
  910, 35, \dodoi{10.3847/1538-4357/abe1bd}

\bibitem[{{Lister} {et~al.}(2018){Lister}, {Aller}, {Aller}, {Hodge}, {Homan},
  {Kovalev}, {Pushkarev}, \& {Savolainen}}]{Lister2018}
{Lister}, M.~L., {Aller}, M.~F., {Aller}, H.~D., {et~al.} 2018, \apjs, 234, 12,
  \dodoi{10.3847/1538-4365/aa9c44}

\bibitem[{{Mart{\'\i}-Vidal} {et~al.}(2021){Mart{\'\i}-Vidal}, {Mus},
  {Janssen}, {de Vicente}, \& {Gonz{\'a}lez}}]{MartiVidal2021}
{Mart{\'\i}-Vidal}, I., {Mus}, A., {Janssen}, M., {de Vicente}, P., \&
  {Gonz{\'a}lez}, J. 2021, \aap, 646, A52, \dodoi{10.1051/0004-6361/202039527}

\bibitem[{{Matthews} {et~al.}(2018){Matthews}, {Crew}, {Doeleman}, {Lacasse},
  {Saez}, {Alef}, {Akiyama}, {Amestica}, {Anderson}, {Barkats}, {Baudry},
  {Brogui{\`e}re}, {Escoffier}, {Fish}, {Greenberg}, {Hecht}, {Hiriart},
  {Hirota}, {Honma}, {Ho}, {Impellizzeri}, {Inoue}, {Kohno}, {Lopez},
  {Mart{\'\i}-Vidal}, {Messias}, {Meyer-Zhao}, {Mora-Klein}, {Nagar},
  {Nishioka}, {Oyama}, {Pankratius}, {Perez}, {Phillips}, {Pradel}, {Rottmann},
  {Roy}, {Ruszczyk}, {Shillue}, {Suzuki}, \& {Treacy}}]{Matthews2018}
{Matthews}, L.~D., {Crew}, G.~B., {Doeleman}, S.~S., {et~al.} 2018, \pasp, 130,
  015002, \dodoi{10.1088/1538-3873/aa9c3d}

\bibitem[{{Napier}(1989)}]{Napier1989}
{Napier}, P.~J. 1989, in Astronomical Society of the Pacific Conference Series,
  Vol.~6, Synthesis Imaging in Radio Astronomy, ed. R.~A. {Perley}, F.~R.
  {Schwab}, \& A.~H. {Bridle}, 39

\bibitem[{Park \& Algaba(2022)}]{PA2022}
Park, J., \& Algaba, J.~C. 2022, Galaxies, 10, \dodoi{10.3390/galaxies10050102}

\bibitem[{{Park} {et~al.}(2021{\natexlab{a}}){Park}, {Asada}, {Nakamura},
  {Kino}, {Pu}, {Hada}, {Kravchenko}, \& {Giroletti}}]{Park2021c}
{Park}, J., {Asada}, K., {Nakamura}, M., {et~al.} 2021{\natexlab{a}}, \apj,
  922, 180, \dodoi{10.3847/1538-4357/ac26bf}

\bibitem[{{Park} {et~al.}(2021{\natexlab{b}}){Park}, {Byun}, {Asada}, \&
  {Yun}}]{Park2021a}
{Park}, J., {Byun}, D.-Y., {Asada}, K., \& {Yun}, Y. 2021{\natexlab{b}}, \apj,
  906, 85, \dodoi{10.3847/1538-4357/abcc6e}

\bibitem[{{Park} {et~al.}(2019{\natexlab{a}}){Park}, {Hada}, {Kino},
  {Nakamura}, {Ro}, \& {Trippe}}]{Park2019a}
{Park}, J., {Hada}, K., {Kino}, M., {et~al.} 2019{\natexlab{a}}, \apj, 871,
  257, \dodoi{10.3847/1538-4357/aaf9a9}

\bibitem[{{Park} {et~al.}(2018){Park}, {Kam}, {Trippe}, {Kang}, {Byun}, {Kim},
  {Algaba}, {Lee}, {Zhao}, {Kino}, {Shin}, {Hada}, {Lee}, {Oh}, {Hodgson}, \&
  {Sohn}}]{Park2018}
{Park}, J., {Kam}, M., {Trippe}, S., {et~al.} 2018, \apj, 860, 112,
  \dodoi{10.3847/1538-4357/aac490}

\bibitem[{{Park} {et~al.}(2019{\natexlab{b}}){Park}, {Hada}, {Kino},
  {Nakamura}, {Hodgson}, {Ro}, {Cui}, {Asada}, {Algaba}, {Sawada-Satoh}, {Lee},
  {Cho}, {Shen}, {Jiang}, {Trippe}, {Niinuma}, {Sohn}, {Jung}, {Zhao},
  {Wajima}, {Tazaki}, {Honma}, {An}, {Akiyama}, {Byun}, {Kim}, {Zhang},
  {Cheng}, {Kobayashi}, {Shibata}, {Lee}, {Roh}, {Oh}, {Yeom}, {Jung}, {Oh},
  {Kim}, {Hwang}, \& {Hagiwara}}]{Park2019b}
{Park}, J., {Hada}, K., {Kino}, M., {et~al.} 2019{\natexlab{b}}, \apj, 887,
  147, \dodoi{10.3847/1538-4357/ab5584}

\bibitem[{{Pesce}(2021)}]{Pesce2021}
{Pesce}, D.~W. 2021, \aj, 161, 178, \dodoi{10.3847/1538-3881/abe3f8}

\bibitem[{{Sault} {et~al.}(1996){Sault}, {Hamaker}, \& {Bregman}}]{Sault1996}
{Sault}, R.~J., {Hamaker}, J.~P., \& {Bregman}, J.~D. 1996, \aaps, 117, 149

\bibitem[{{Shepherd}(1997)}]{Shepherd1997}
{Shepherd}, M.~C. 1997, in Astronomical Society of the Pacific Conference
  Series, Vol. 125, Astronomical Data Analysis Software and Systems VI, ed.
  G.~{Hunt} \& H.~{Payne}, 77

\bibitem[{{Smirnov}(2011)}]{Smirnov2011}
{Smirnov}, O.~M. 2011, \aap, 527, A106, \dodoi{10.1051/0004-6361/201016082}

\bibitem[{{Thum} {et~al.}(2008){Thum}, {Wiesemeyer}, {Paubert}, {Navarro}, \&
  {Morris}}]{Thum2008}
{Thum}, C., {Wiesemeyer}, H., {Paubert}, G., {Navarro}, S., \& {Morris}, D.
  2008, \pasp, 120, 777, \dodoi{10.1086/590190}

\bibitem[{{Tiede}(2022)}]{Tiede2022}
{Tiede}, P. 2022, The Journal of Open Source Software, 7, 4457,
  \dodoi{10.21105/joss.04457}

\bibitem[{{Virtanen} {et~al.}(2020){Virtanen}, {Gommers}, {Oliphant},
  {Haberland}, {Reddy}, {Cournapeau}, {Burovski}, {Peterson}, {Weckesser},
  {Bright}, {van der Walt}, {Brett}, {Wilson}, {Jarrod Millman}, {Mayorov},
  {Nelson}, {Jones}, {Kern}, {Larson}, {Carey}, {Polat}, {Feng}, {Moore}, {Vand
  erPlas}, {Laxalde}, {Perktold}, {Cimrman}, {Henriksen}, {Quintero}, {Harris},
  {Archibald}, {Ribeiro}, {Pedregosa}, {van Mulbregt}, \&
  {Contributors}}]{Scipy2020}
{Virtanen}, P., {Gommers}, R., {Oliphant}, T.~E., {et~al.} 2020, Nature
  Methods, 17, 261, \dodoi{https://doi.org/10.1038/s41592-019-0686-2}

\bibitem[{{Wardle} \& {Kronberg}(1974)}]{WK1974}
{Wardle}, J.~F.~C., \& {Kronberg}, P.~P. 1974, \apj, 194, 249,
  \dodoi{10.1086/153240}

\bibitem[{{Zhao} {et~al.}(2022){Zhao}, {Gomez}, {Fuentes}, {Krichbaum},
  {Traianou}, {Lico}, {Cho}, {Ros}, {Komossa}, {Akiyama}, {Asada}, {Blackburn},
  {Britzen}, {Bruni}, {Crew}, {Dahale}, {Dey}, {Gold}, {Gopakumar}, {Issaoun},
  {Janssen}, {Jorstad}, {Kim}, {Koay}, {Kovalev}, {Koyama}, {Lobanov},
  {Loinard}, {Lu}, {Markoff}, {Marscher}, {Marti-Vidal}, {Mizuno}, {Park},
  {Savolainen}, \& {Toscano}}]{Zhao2022}
{Zhao}, G.-Y., {Gomez}, J.~L., {Fuentes}, A., {et~al.} 2022, arXiv e-prints,
  arXiv:2205.00554.
\newblock \doarXiv{2205.00554}

\end{thebibliography}
\bibliographystyle{aasjournal}

\end{document}